\documentclass[reprint,showpacs,preprintnumbers,nofootinbib,amsmath,amssymb,aps,prd,floatfix]{revtex4-1}
\pdfoutput=1
\usepackage{graphicx}
\usepackage{bm}
\usepackage{subfigure}
\usepackage{url}
\usepackage[hyperindex]{hyperref}
\usepackage{color}
\usepackage[ddmmyy,24hr]{datetime}
\usepackage{bigdelim}
\usepackage{booktabs}
\usepackage{dcolumn}
\usepackage{multirow}
\usepackage{subfigure}
\usepackage{cancel}
\usepackage{stackrel}
\usepackage{paralist}
\usepackage{xspace}
\newcommand{\nua}[1]{\ensuremath{\rlap{\kern-2.5pt\ensuremath{\overset{\scriptscriptstyle(-)}{\phantom{\nu}}}}{\ensuremath{{\nu}_{#1}}}}}

\usepackage{enumitem}

\makeatletter
\def\namedlabel#1#2{\begingroup
    #2%
    \def\@currentlabel{#2}%
    \phantomsection\label{#1}\endgroup
}
\makeatother
\begin{document}

\title{Reactor Fuel Fraction Information on the Antineutrino Anomaly}

\author{C. Giunti}
%\email{giunti@to.infn.it}
%\altaffiliation[Also at the ]{Department of Theoretical Physics, University of Torino, Italy}
\affiliation{INFN, Sezione di Torino, Via P. Giuria 1, I--10125 Torino, Italy}

\author{X.P. Ji}
%\email{jixp@mail.tsinghua.edu.cn}
\affiliation{Department of Engineering Physics, Tsinghua University, Beijing 100084, China}

\author{M. Laveder}
%\email{laveder@pd.infn.it}
\affiliation{Dipartimento di Fisica e Astronomia ``G. Galilei'', Universit\`a di Padova,
and
INFN, Sezione di Padova,
Via F. Marzolo 8, I--35131 Padova, Italy}

\author{Y.F. Li}
%\email{liyufeng@ihep.ac.cn}
\affiliation{Institute of High Energy Physics,
Chinese Academy of Sciences,
and
School of Physical Sciences, University of Chinese Academy of Sciences, Beijing 100049, China}

\author{B.R. Littlejohn}
%\email{blittlej@iit.edu}
\affiliation{Illinois Institute of Technology, Chicago, IL 60616, USA}

%\date{\dayofweekname{\day}{\month}{\year} \ddmmyydate\today, \currenttime}
\date{3 August 2017}

\begin{abstract}
We analyzed
the evolution data of the Daya Bay reactor neutrino experiment
in terms of short-baseline active-sterile neutrino oscillations
taking into account the theoretical uncertainties of the reactor antineutrino fluxes.
We found that oscillations are disfavored at
$2.6\sigma$
with respect
to a suppression of the ${}^{235}\text{U}$
reactor antineutrino flux
and at
$2.5\sigma$
with respect to variations of the
${}^{235}\text{U}$ and ${}^{239}\text{Pu}$
fluxes.
On the other hand,
the analysis of
the rates of the short-baseline reactor neutrino experiments
favor active-sterile neutrino oscillations
and disfavor
the suppression of the
${}^{235}\text{U}$
flux at
$3.1\sigma$
and
variations of the
${}^{235}\text{U}$ and ${}^{239}\text{Pu}$
fluxes at
$2.8\sigma$.
We also found that both the Daya Bay evolution data and the global rate data are well-fitted with composite hypotheses including variations of the
${}^{235}\text{U}$ or ${}^{239}\text{Pu}$
fluxes in addition to active-sterile neutrino oscillations.
A combined analysis of the Daya Bay evolution data and the global rate data
shows a slight preference for oscillations
with respect to variations of the
${}^{235}\text{U}$ and ${}^{239}\text{Pu}$
fluxes.
However,
the best fits of the combined data are given by the composite models,
with a preference for the model with
an enhancement of the
${}^{239}\text{Pu}$
flux
and relatively large oscillations.
\end{abstract}


\maketitle

\section{Introduction}
\label{sec:intro}

The Daya Bay collaboration
presented recently~\cite{An:2017osx}
the results of the measurement of the correlation between the
reactor fuel evolution and
the changes in the antineutrino detection rate which is quantified by the cross section per fission
$\sigma_{f}$,
given by
\begin{equation}
\sigma_{f}
=
\sum_{i} F_{i} \sigma_{f,i}
,
\label{csf}
\end{equation}
where
$F_{i}^{a}$ and $\sigma_{f,i}$
are the effective fission fractions and the cross sections per fission
of the four fissionable isotopes
${}^{235}\text{U}$,
${}^{238}\text{U}$,
${}^{239}\text{Pu}$,
${}^{241}\text{Pu}$,
denoted, respectively, with the label
$i=235,238,239,241$.

The Daya Bay collaboration presented in Fig.~2 of Ref.~\cite{An:2017osx}
the values of $\sigma_{f}$
for eight values of the effective ${}^{239}\text{Pu}$ fission fraction $F_{239}$.
They fitted these data
allowing variations of the two main cross sections per fission
$\sigma_{f,235}$
and
$\sigma_{f,239}$,
with the assumption that
$\sigma_{f,238}$
and
$\sigma_{f,241}$
have the Saclay+Huber theoretical values~\cite{Mueller:2011nm,Mention:2011rk,Huber:2011wv}
with enlarged 10\% uncertainties.
They also compared the best-fit of this analysis
with the best-fit obtained under the hypothesis of active-sterile neutrino oscillations,
which predicts the same suppression for the four cross sections per fission
with respect to their theoretical value.
They obtained $\Delta\chi^2/\text{NDF}=7.9/1$,
corresponding to a $p$-value of 0.49\%,
which disfavors the active-sterile oscillations hypothesis by $2.8\sigma$.
In this calculation
the uncertainties of the theoretical
calculation of the four cross sections per fission
were not taken into account.

In this paper we present the results of analyses of the Daya Bay evolution data
\cite{An:2017osx}
with least-squares functions that take into account explicitly
the uncertainties of the theoretical
calculation of the four cross sections per fission.
Moreover,
we consider additional models with independent variations
of the
${}^{235}\text{U}$ and ${}^{239}\text{Pu}$
fluxes
with and without active-sterile neutrino oscillations,
and
we extend the analysis taking into account also the
information on the cross sections per fission of all the other reactor antineutrino experiments
which have different fuel fractions.
We also perform proper statistical comparisons of the non-nested models under consideration
through Monte Carlo estimations of the $p$-values.

Given a set of data labeled with the index $a$ on the cross section per fission
for different values of the fuel fractions,
we write the theoretical predictions as
\begin{equation}
\sigma_{f,a}^{\text{th}}
=
\sum_{i} F_{i}^{a} r_{i} \sigma_{f,i}^{\text{SH}}
,
\label{csfth}
\end{equation}
where $i=235,238,239,241$ and
$\sigma_{f,i}^{\text{SH}}$
are the Saclay+Huber cross sections per fission.
The coefficients $r_{i}$ are introduced in order to take into account the uncertainties of
the Saclay+Huber cross sections per fission
or to study independent variations of the antineutrino fluxes
from the four fissionable isotopes
with respect to
the Saclay+Huber theoretical values~\cite{Mueller:2011nm,Mention:2011rk,Huber:2011wv}.

We consider the following models:

\begin{description}

\item[\namedlabel{case:235}{235}]
A variation of the cross section per fission
of the antineutrino flux from
${}^{235}\text{U}$ only.

In this case,
we analyze the data with the least-squares statistic
\begin{align}
\chi^2
=
\null & \null
\sum_{a,b}
\left( \sigma_{f,a}^{\text{th}} - \sigma_{f,a}^{\text{exp}} \right)
(V_{\text{exp}}^{-1})_{ab}
\left( \sigma_{f,b}^{\text{th}} - \sigma_{f,b}^{\text{exp}} \right)
\nonumber
\\
\null & \null
+ \sum_{i,j=238,239,241}
\left( r_{i} - 1 \right)
(V_{\text{SH}}^{-1})_{ij}
\left( r_{j} - 1 \right)
,
\label{chi:235}
\end{align}
where
$\sigma_{f,a}^{\text{exp}}$
are the measured cross sections per fission,
$V_{\text{exp}}$
is the experimental covariance matrix,
and
$V_{\text{SH}}$
is the covariance matrix of the fractional uncertainties of the Saclay-Huber theoretical calculation
of the antineutrino fluxes
from the four fissionable isotopes
(given in Table~3 of Ref.~\cite{Gariazzo:2017fdh}).

In this analysis there is only one parameter determined by the fit:
$r_{235}$.
The parameters
$r_{238}$,
$r_{239}$, and
$r_{241}$
are nuisance parameters.

\item[\namedlabel{case:235+239}{235+239}]
Independent variations of the cross sections per fission
of the antineutrino fluxes from
${}^{235}\text{U}$
and
${}^{239}\text{Pu}$.

In this case,
we analyze the data with the least-squares statistic
\begin{align}
\chi^2
=
\null & \null
\sum_{a,b}
\left( \sigma_{f,a}^{\text{th}} - \sigma_{f,a}^{\text{exp}} \right)
(V_{\text{exp}}^{-1})_{ab}
\left( \sigma_{f,b}^{\text{th}} - \sigma_{f,b}^{\text{exp}} \right)
\nonumber
\\
\null & \null
+ \sum_{i,j=238,241}
\left( r_{i} - 1 \right)
(V_{\text{SH}}^{-1})_{ij}
\left( r_{j} - 1 \right)
.
\label{chi:235+239}
\end{align}

In this analysis there are two parameters determined by the fit:
$r_{235}$ and $r_{239}$.
The parameters
$r_{238}$ and $r_{241}$
are nuisance parameters.

\item[\namedlabel{case:OSC}{OSC}]
Active-sterile neutrino oscillations, in which the measured cross sections per fission
are suppressed with respect to the theoretical cross sections per fission
$\sigma_{f,a}^{\text{th}}$
by the survival probability
$P_{ee}$
which is independent of the ${}^{239}\text{Pu}$ fraction $F_{239}$.

In this case,
we analyze the data with the least-squares statistic
\begin{align}
\chi^2
=
\null & \null
\sum_{a,b}
\left( P_{ee} \sigma_{f,a}^{\text{th}} - \sigma_{f,a}^{\text{exp}} \right)
(V_{\text{exp}}^{-1})_{ab}
\left( P_{ee} \sigma_{f,b}^{\text{th}} - \sigma_{f,b}^{\text{exp}} \right)
\nonumber
\\
\null & \null
+ \sum_{i,j}
\left( r_{i} - 1 \right)
(V_{\text{SH}}^{-1})_{ij}
\left( r_{j} - 1 \right)
.
\label{chi:OSC}
\end{align}

In the analysis of the Daya Bay evolution data
there is only one parameter determined by the fit:
$P_{ee}$.
The parameters
$r_{235}$,
$r_{238}$,
$r_{239}$, and
$r_{241}$
are nuisance parameters.
In the analysis of the other reactor antineutrino data
we take into account that $P_{ee}$
depends on the neutrino mixing parameters
$\Delta{m}^2_{41}$
and
$\sin^22\vartheta_{ee}$
in the simplest 3+1 active-sterile neutrino mixing model
(see Ref.~\cite{Gariazzo:2015rra}).
Hence, in this case there are two parameters determined by the fit:
$\Delta{m}^2_{41}$
and
$\sin^22\vartheta_{ee}$.

\item[\namedlabel{case:235+OSC}{235+OSC}]
A variation of the cross section per fission
of the antineutrino flux from
${}^{235}\text{U}$
and
active-sterile neutrino oscillations with a survival probability
$P_{ee}$ as in the \textbf{\ref{case:OSC}} model.

In this case,
we analyze the data with the least-squares statistic
\begin{align}
\chi^2
=
\null & \null
\sum_{a,b}
\left( P_{ee} \sigma_{f,a}^{\text{th}} - \sigma_{f,a}^{\text{exp}} \right)
(V_{\text{exp}}^{-1})_{ab}
\left( P_{ee} \sigma_{f,b}^{\text{th}} - \sigma_{f,b}^{\text{exp}} \right)
\nonumber
\\
\null & \null
+ \sum_{i,j=238,239,241}
\left( r_{i} - 1 \right)
(V_{\text{SH}}^{-1})_{ij}
\left( r_{j} - 1 \right)
.
\label{chi:235+OSC}
\end{align}

In the analysis of the Daya Bay evolution data
there are two parameters determined by the fit:
$r_{235}$ and $P_{ee}$.
The parameters
$r_{238}$ and $r_{241}$
are nuisance parameters.
In the analysis of the other reactor antineutrino data
we take into account that $P_{ee}$
depends on
$\Delta{m}^2_{41}$
and
$\sin^22\vartheta_{ee}$
as in the \textbf{\ref{case:OSC}} model.
Therefore, in this case there are three parameters determined by the fit:
$r_{235}$,
$\Delta{m}^2_{41}$,
and
$\sin^22\vartheta_{ee}$.

\item[\namedlabel{case:239+OSC}{239+OSC}]
This model is similar to the \textbf{\ref{case:235+OSC}} model,
with
${}^{235}\text{U} \leftrightarrows {}^{239}\text{Pu}$.
The number of parameters determined by the fit
is two in the analysis of the Daya Bay evolution data
($r_{239}$ and $P_{ee}$)
and
three in the analysis of the other reactor antineutrino data
($r_{239}$,
$\Delta{m}^2_{41}$,
and
$\sin^22\vartheta_{ee}$).

\end{description}

In Section~\ref{sec:dayabay} we analyze the Daya Bay evolution data,
in Section~\ref{sec:rea} we analyze
the reactor antineutrino data which were available
before the release of the Daya Bay fuel evolution data in Ref.~\cite{An:2017osx},
and
in Section~\ref{sec:all} we perform the combined analysis.

\section{Daya Bay evolution}
\label{sec:dayabay}

\begin{table}[!b]
\centering
\begin{tabular}[t]{c}
\\
\hline
$\chi^{2}_{\text{min}}$\\
NDF\\
GoF\\
$P_{ee}$\\
$r_{235}$\\
$r_{239}$\\
%\hline%
\end{tabular}%
\begin{tabular}[t]{c}
\textbf{\ref{case:235}}\\
\hline
$3.8$\\
$7$\\
$80\%$\\
$-$\\
$0.927$\\
$-$\\
%\hline%
\end{tabular}%
\begin{tabular}[t]{c}
\textbf{\ref{case:235+239}}\\
\hline
$3.6$\\
$6$\\
$73\%$\\
$-$\\
$0.922$\\
$0.974$\\
%\hline%
\end{tabular}%
\begin{tabular}[t]{c}
\textbf{\ref{case:OSC}}\\
\hline
$9.5$\\
$7$\\
$22\%$\\
$0.942$\\
$-$\\
$-$\\
%\hline%
\end{tabular}%
\begin{tabular}[t]{c}
\textbf{\ref{case:235+OSC}}\\
\hline
$3.6$\\
$6$\\
$72\%$\\
$0.984$\\
$0.937$\\
$-$\\
%\hline%
\end{tabular}%
\begin{tabular}[t]{c}
\textbf{\ref{case:239+OSC}}\\
\hline
$3.8$\\
$6$\\
$71\%$\\
$0.928$\\
$-$\\
$1.094$\\
%\hline%
\end{tabular}%
\caption{ \label{tab:dby}
Fits of the Daya Bay evolution data~\cite{An:2017osx}.
}
\end{table}

\begin{figure}[!t]
\centering
\includegraphics*[width=\linewidth]{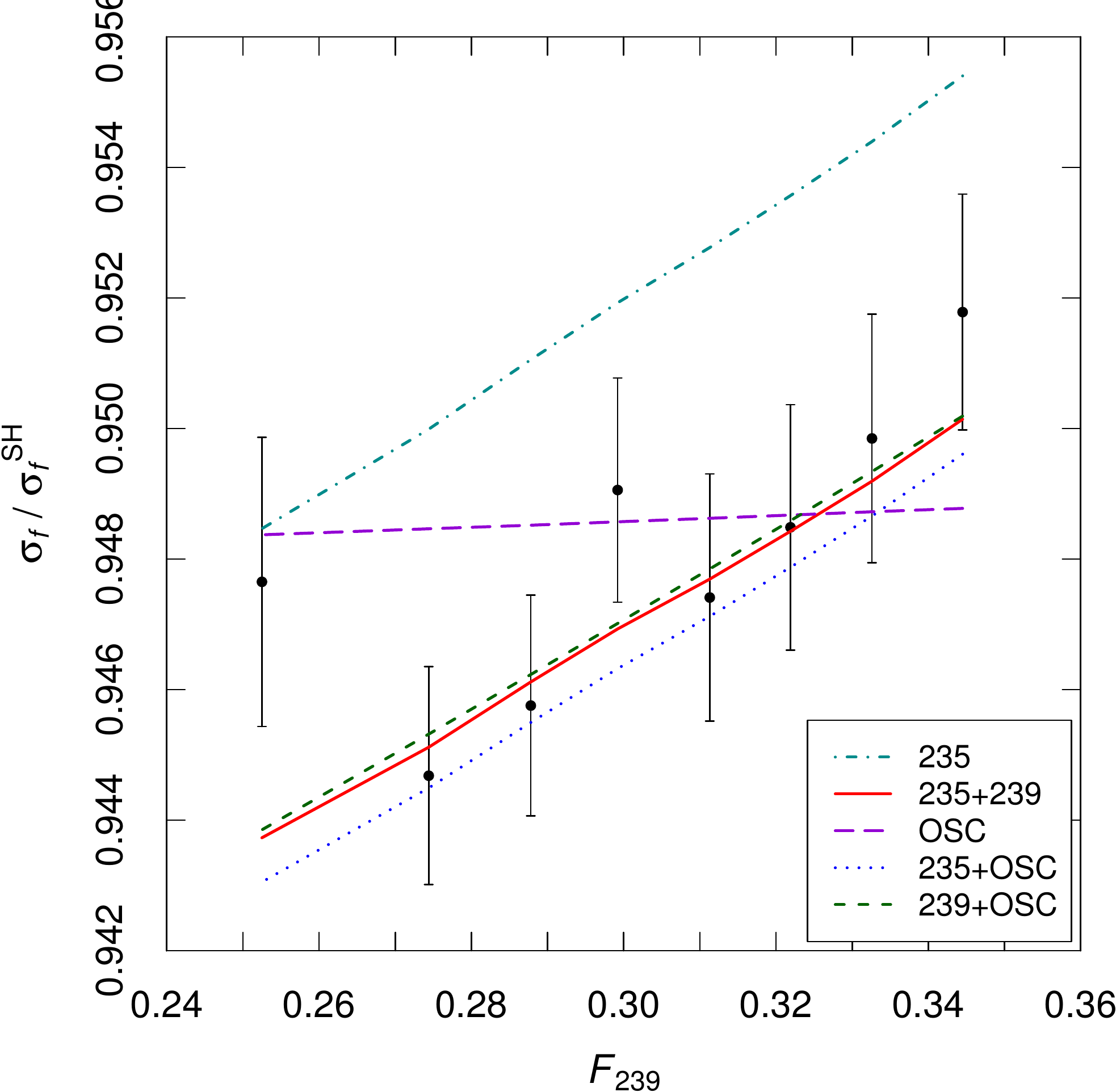}
\caption{ \label{fig:dby-evo2}
Fits of the Daya Bay evolution data~\cite{An:2017osx}
normalized to the Saclay-Huber theoretical predictions~\cite{Mueller:2011nm,Mention:2011rk,Huber:2011wv}.
The error bars show only the uncorrelated statistical uncertainties.
}
\end{figure}

The results of the different fits of the Daya Bay evolution data
are given in Tab.~\ref{tab:dby}
where we list the values of the minimum $\chi^2$,
the number of degrees of freedom and the goodness-of-fit.
In Tab.~\ref{tab:dby} we also list the best-fit values of the fitted parameters.

Figure \ref{fig:dby-evo2} shows the comparison of the different fits
with the Daya Bay evolution data normalized to the Saclay-Huber theoretical
cross sections per fission
\cite{Mueller:2011nm,Mention:2011rk,Huber:2011wv}.
Note that the Daya Bay evolution data have the following two important features:

\begin{description}

\item[\namedlabel{F1}{F1}]
A suppression of $\sigma_{f}$ with respect to $\sigma_{f}^{\text{SH}}$ in agreement with the reactor antineutrino anomaly.
This feature can be fitted with at least one of the $r_{i}$ and $P_{ee}$
smaller than one (if the others are equal to one).

\item[\namedlabel{F2}{F2}]
An increase of
$\sigma_{f} / \sigma_{f}^{\text{SH}}$
with
$F_{239}$.
This feature can be fitted if
\begin{equation}
\frac{d}{d F_{239}}
\,
\frac{\sigma_{f,a}^{\text{th}}}{\sigma_{f,a}^{\text{SH}}}
>
0
,
\label{drv1}
\end{equation}
where
\begin{equation}
\sigma_{f,a}^{\text{SH}}
=
\sum_{i} F_{i}^{a} \sigma_{f,i}^{\text{SH}}
.
\label{drv2}
\end{equation}
The inequality (\ref{drv1}) is satisfied for
\begin{equation}
\sum_{i}
\frac{d F_{i}^{a}}{d F_{239}}
\,
r_{i}
\sigma_{f,i}^{\text{SH}}
>
\frac{\sigma_{f,a}^{\text{th}}}{\sigma_{f,a}^{\text{SH}}}
\,
\frac{d \sigma_{f,a}^{\text{SH}}}{d F_{239}}
,
\label{drv3}
\end{equation}
with
\begin{equation}
\frac{d \sigma_{f,a}^{\text{SH}}}{d F_{239}}
\simeq
-2.4
<
0
.
\label{drv4}
\end{equation}

\end{description}

From Tab.~\ref{tab:dby} one can see that all the fits have acceptable goodness-of-fit,
but the \textbf{\ref{case:OSC}} fit
corresponding to active-sterile oscillations
has a goodness-of-fit which is significantly lower than the others,
because it corresponds to a constant
$\sigma_{f} / \sigma_{f}^{\text{SH}}$
and cannot
fit feature~\textbf{\ref{F2}}.

The results of our analysis agree with the conclusion of the Daya Bay collaboration~\cite{An:2017osx}
that the \textbf{\ref{case:235}} model fits well the data
and little is gained by allowing also the variation of
$\sigma_{f,239}$ in the
\textbf{\ref{case:235+239}} model.
The shift in Fig.~\ref{fig:dby-evo2}
of the line corresponding to the \textbf{\ref{case:235}} model
with respect to an ideal line fitting the data by eye
is allowed by the large correlated systematic uncertainties
of the Daya Bay bins \cite{An:2017osx}.

The excellent fit in the \textbf{\ref{case:235}} model is due to the
fact that it can fit the two features of the Daya Bay evolution data listed above.
It can obviously fit feature~\textbf{\ref{F1}} with $r_{235}<1$.
It can also fit feature~\textbf{\ref{F2}},
because for $r_{235}<1$ and $r_{238}=r_{239}=r_{241}=1$
the condition (\ref{drv3}) becomes
\begin{equation}
-
\frac{d F_{235}^{a}}{d F_{239}}
>
-
\frac{d \sigma_{f,a}^{\text{SH}}}{d F_{239}}
\,
\frac{F_{235}^{a}}{\sigma_{f,a}^{\text{SH}}}
.
\label{drv235}
\end{equation}
This condition is satisfied,
because numerically the left-hand side is about
1.30
and the right-hand side is between
0.20
and
0.24.

Obviously,
the
\textbf{\ref{case:235+OSC}} model can provide a fit which is at least as good as the
\textbf{\ref{case:235}} model,
with the additional possibility to improve the fit of
feature~\textbf{\ref{F1}}
with $P_{ee}<1$.

It is maybe more surprising that also the \textbf{\ref{case:239+OSC}} model
fits better than the \textbf{\ref{case:235}} model for $r_{239}>1$.
This can happen because the condition (\ref{drv3}) for fitting feature~\textbf{\ref{F2}} is always satisfied for
$r_{239}>1$ and $r_{235}=r_{238}=r_{241}=1$.
Then, a sufficiently small value of $P_{ee}<1$
allows us to fit feature~\textbf{\ref{F1}}
in spite of the increase of $\sigma_{f,a}^{\text{th}}$ due to
$r_{239}>1$.

Nested models can be compared in the frequentist approach by
calculating the $p$-value of the $\chi^2_{\text{min}}$ difference,
which has a $\chi^2$ distribution corresponding to the difference of the number of degrees of freedom
of the two models.
With this method we can compare only the nested models
\textbf{\ref{case:235}} and \textbf{\ref{case:235+OSC}},
because the $\chi^2$ in Eq.~(\ref{chi:235}) can be obtained from that in Eq.~(\ref{chi:235+OSC})
with the constraint $P_{ee}=1$.
In this comparison, we have
$\Delta\chi^2 = 0.2$
with one degree of freedom.
Hence,
the null hypothesis \textbf{\ref{case:235}}
cannot be rejected in favor of the alternative more complex hypothesis
\textbf{\ref{case:235+OSC}}.

Also non-nested models can be compared considering the $\chi^2_{\text{min}}$ difference,
but one must calculate the $p$-value with a Monte Carlo.
In this case one must consider as the null hypothesis the model which has the higher
$\chi^2_{\text{min}}$
and
generate many sets of synthetic data assuming the null hypothesis.
The fits of all the
sets of synthetic data with the two models under consideration
gives the distribution of the $\chi^2_{\text{min}}$ difference
from which one can calculate the
$p$-value
of the observed $\chi^2_{\text{min}}$ difference.

We do not bother to consider the comparison of the
\textbf{\ref{case:235}}
and
\textbf{\ref{case:235+239}}
models,
since the small
$\Delta\chi^2_{\text{min}} = 0.2$
cannot lead to the rejection of the null hypothesis \textbf{\ref{case:235}}.

On the other hand,
it is interesting to compare the
\textbf{\ref{case:OSC}}
and
\textbf{\ref{case:235}}
models which have
$\Delta\chi^2_{\text{min}} = 5.7$.
According to our Monte Carlo simulation,
the $p$-value of the null hypothesis \textbf{\ref{case:OSC}}
is
$0.85\%$.
Hence,
the comparison of the
\textbf{\ref{case:OSC}}
and
\textbf{\ref{case:235}}
models disfavors the
\textbf{\ref{case:OSC}} model at the
$2.6\sigma$
level.

We also compared with a Monte Carlo the
\textbf{\ref{case:OSC}}
and
\textbf{\ref{case:235+239}}
models which have
$\Delta\chi^2_{\text{min}} = 5.9$.
We found that
the null hypothesis \textbf{\ref{case:OSC}}
has a $p$-value of 
$1.3\%$,
which is larger than in the previous case because
the \textbf{\ref{case:235+239}} model
has one parameter more than the
\textbf{\ref{case:235}}
model.
Thus,
in this case,
the \textbf{\ref{case:OSC}} model
is disfavored at the
$2.5\sigma$
level,
which is slightly less stringent
than the $2.8\sigma$ obtained by the Daya Bay collaboration~\cite{An:2017osx}
without considering the theoretical uncertainties.

\begin{figure}[!t]
\centering
\includegraphics*[width=\linewidth]{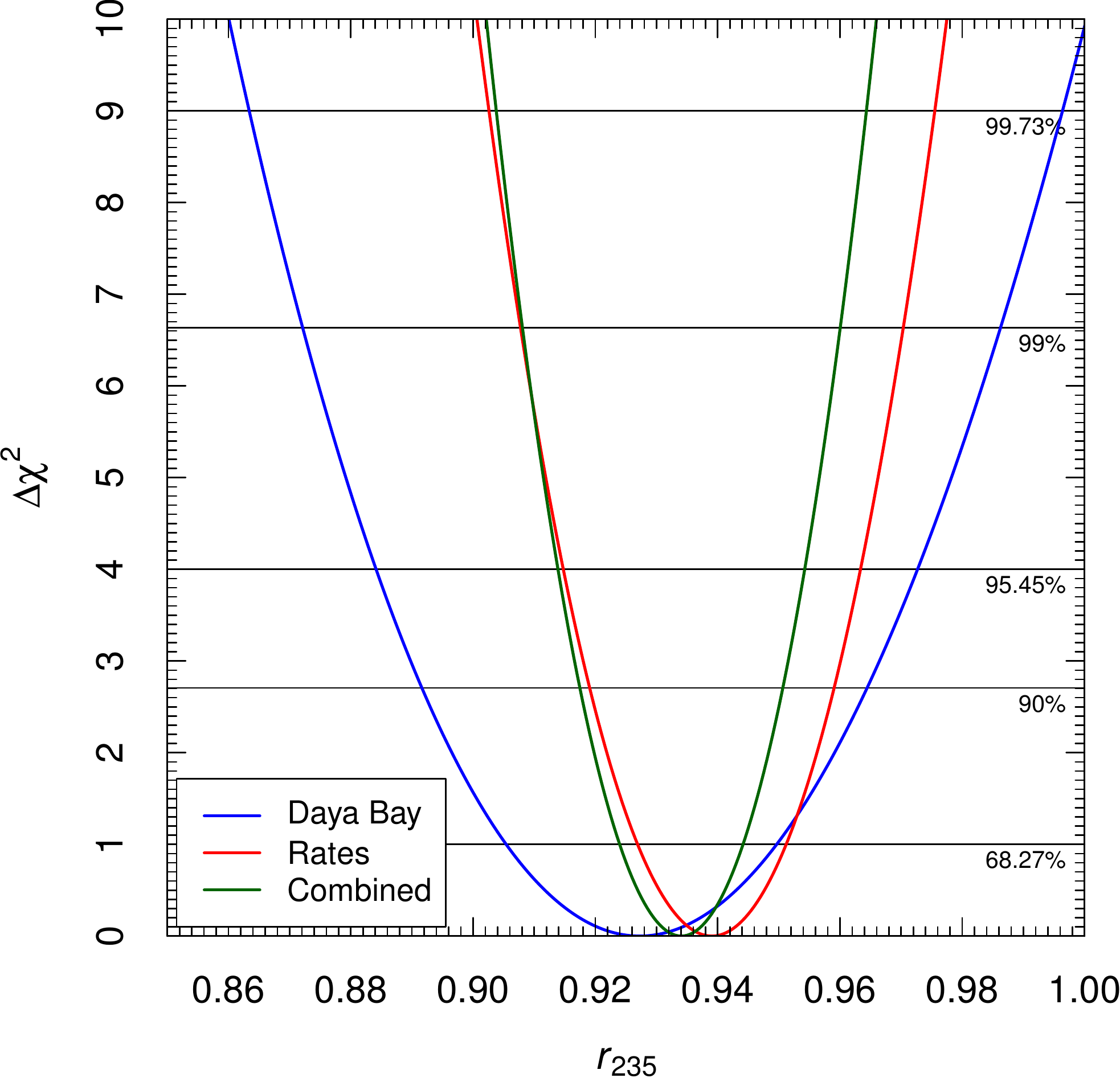}
\caption{ \label{fig:235}
Marginal $\Delta\chi^2 = \chi^2 - \chi^2_{\text{min}}$
for the factor $r_{235}$
obtained
from the fit of the Daya Bay evolution data~\cite{An:2017osx} (Daya Bay),
from the fit of the reactor rates (Rates),
and from the combined fit (Combined)
with the \textbf{\ref{case:235}} model.
}
\end{figure}

\begin{figure}[!t]
\centering
\includegraphics*[width=\linewidth]{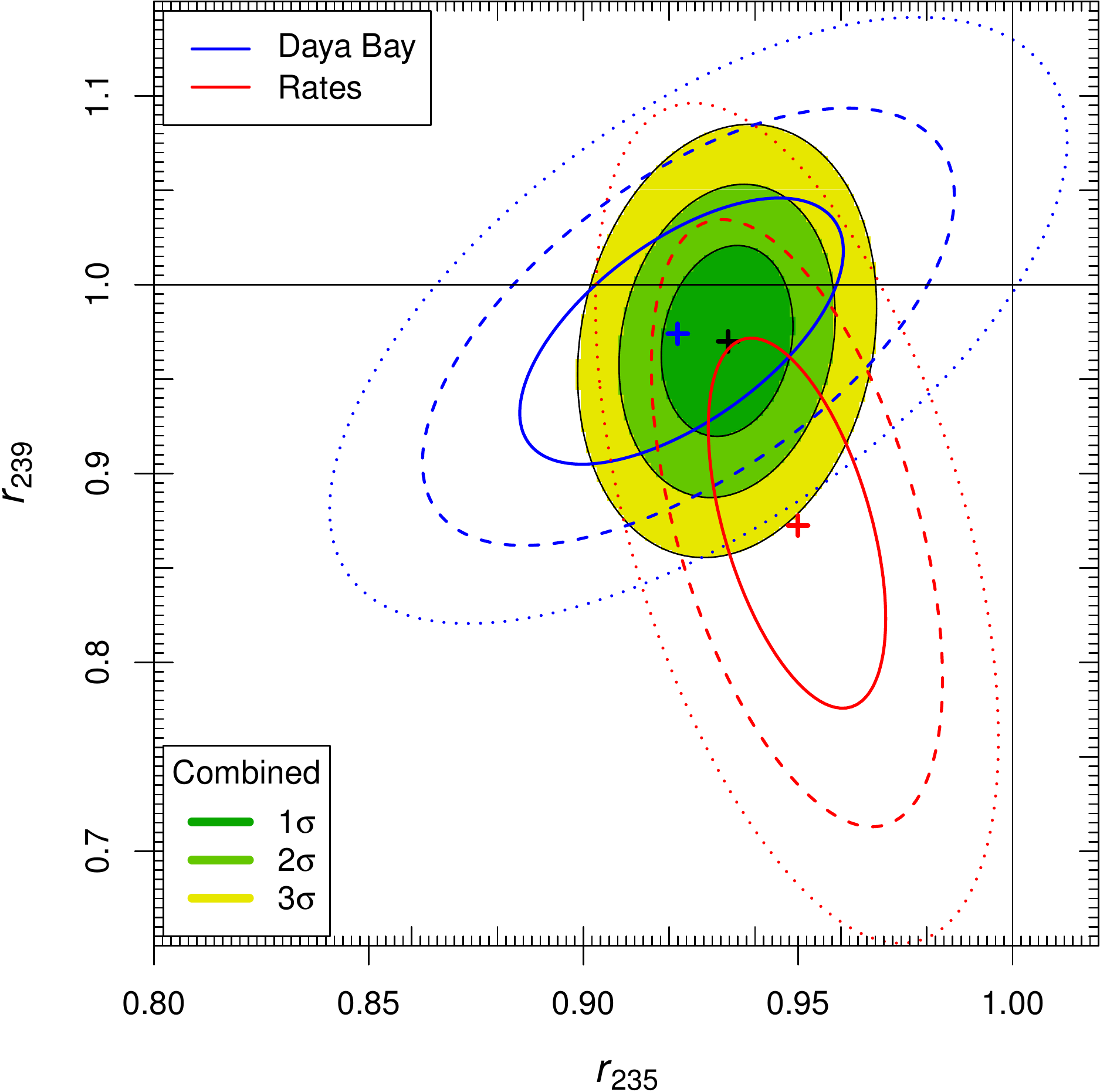}
\caption{ \label{fig:235+239}
Allowed regions in the $r_{235}$--$r_{239}$ plane
obtained
from the fit of the Daya Bay evolution data~\cite{An:2017osx} (Daya Bay),
from the fit of the reactor rates (Rates),
and from the combined fit (Combined)
with the \textbf{\ref{case:235+239}} model.
The best fit points are indicated by crosses.
For the Daya Bay and Rates fits the
$1\sigma$,
$2\sigma$, and
$3\sigma$
allowed regions are limited,
respectively,
by solid, dashed, and dotted lines.
}
\end{figure}

\begin{figure}[!t]
\centering
\includegraphics*[width=\linewidth]{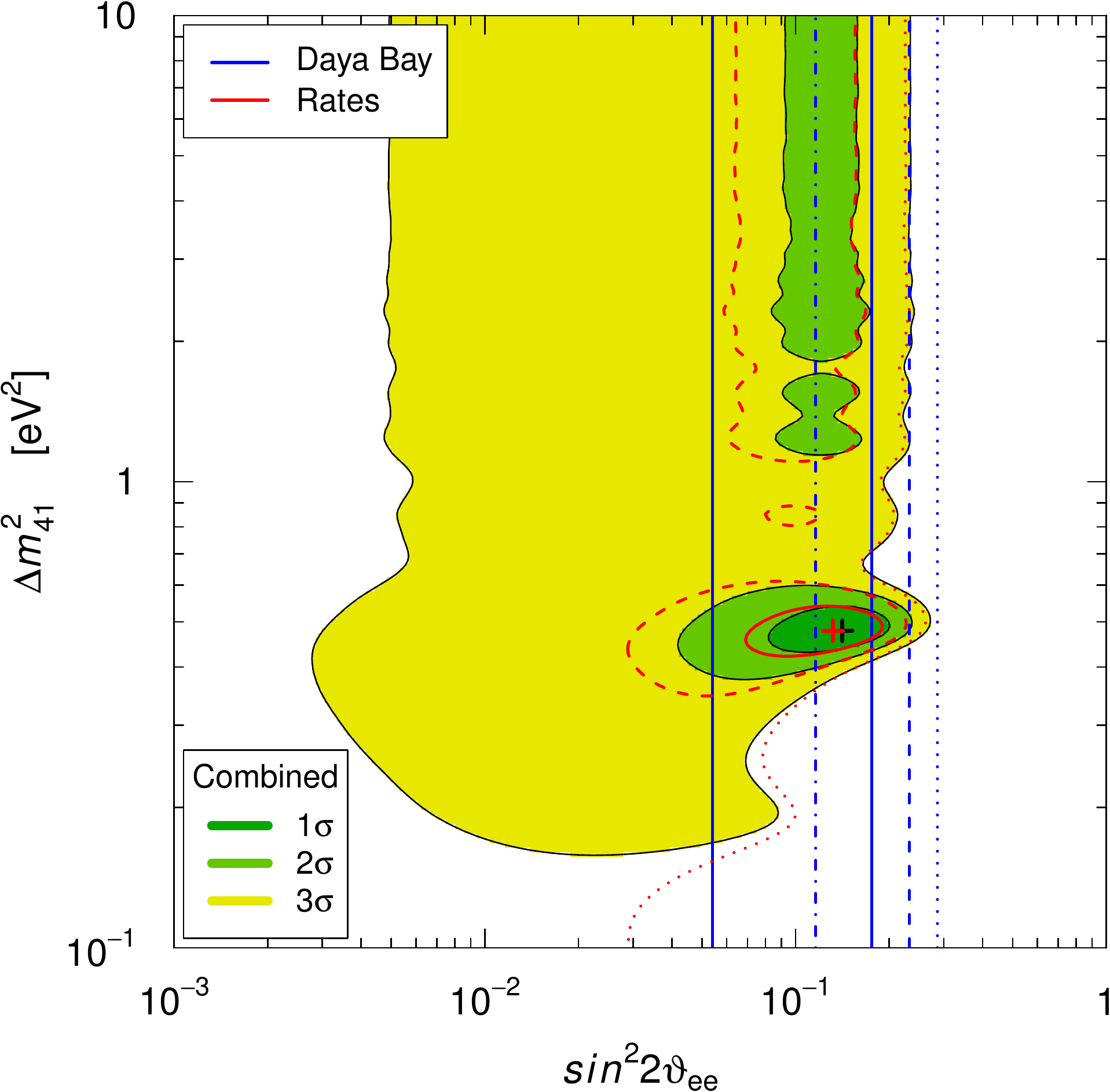}
\caption{ \label{fig:OSC}
Allowed regions in the $\sin^22\vartheta_{ee}$--$\Delta{m}^2_{41}$ plane
obtained
from the fit of the Daya Bay evolution data~\cite{An:2017osx} (Daya Bay),
from the fit of the reactor rates (Rates),
and from the combined fit (Combined)
with the \textbf{\ref{case:OSC}} model.
The best fit points are indicated by crosses,
except for the fit of the Daya Bay evolution data for which
the best fit is the vertical dash-dotted line.
For the Daya Bay and Rates fits the
$1\sigma$,
$2\sigma$, and
$3\sigma$
allowed regions are limited,
respectively,
by solid, dashed, and dotted lines.
}
\end{figure}

\begin{figure}[!t]
\centering
\includegraphics*[width=\linewidth]{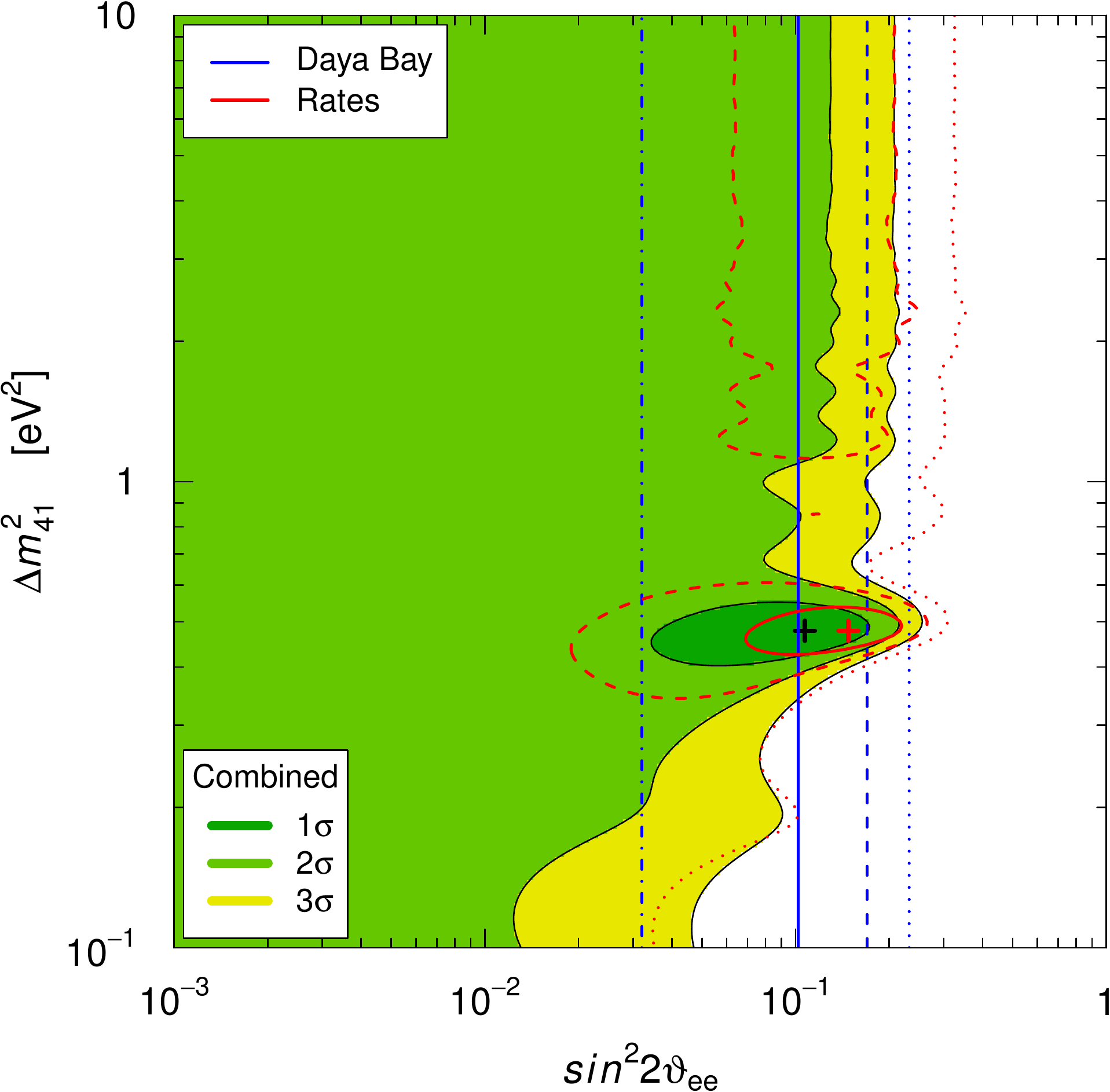}
\caption{ \label{fig:235+OSC}
Allowed regions in the $\sin^22\vartheta_{ee}$--$\Delta{m}^2_{41}$ plane
obtained
from the fit of the Daya Bay evolution data~\cite{An:2017osx} (Daya Bay),
from the fit of the reactor rates (Rates),
and from the combined fit (Combined)
with the \textbf{\ref{case:235+OSC}} model.
The best fit points are indicated by crosses,
except for the fit of the Daya Bay evolution data for which
the best fit is the vertical dash-dotted line.
For the Daya Bay and Rates fits the
$1\sigma$,
$2\sigma$, and
$3\sigma$
allowed regions are limited,
respectively,
by solid, dashed, and dotted lines.
}
\end{figure}

\begin{figure}[!t]
\centering
\includegraphics*[width=\linewidth]{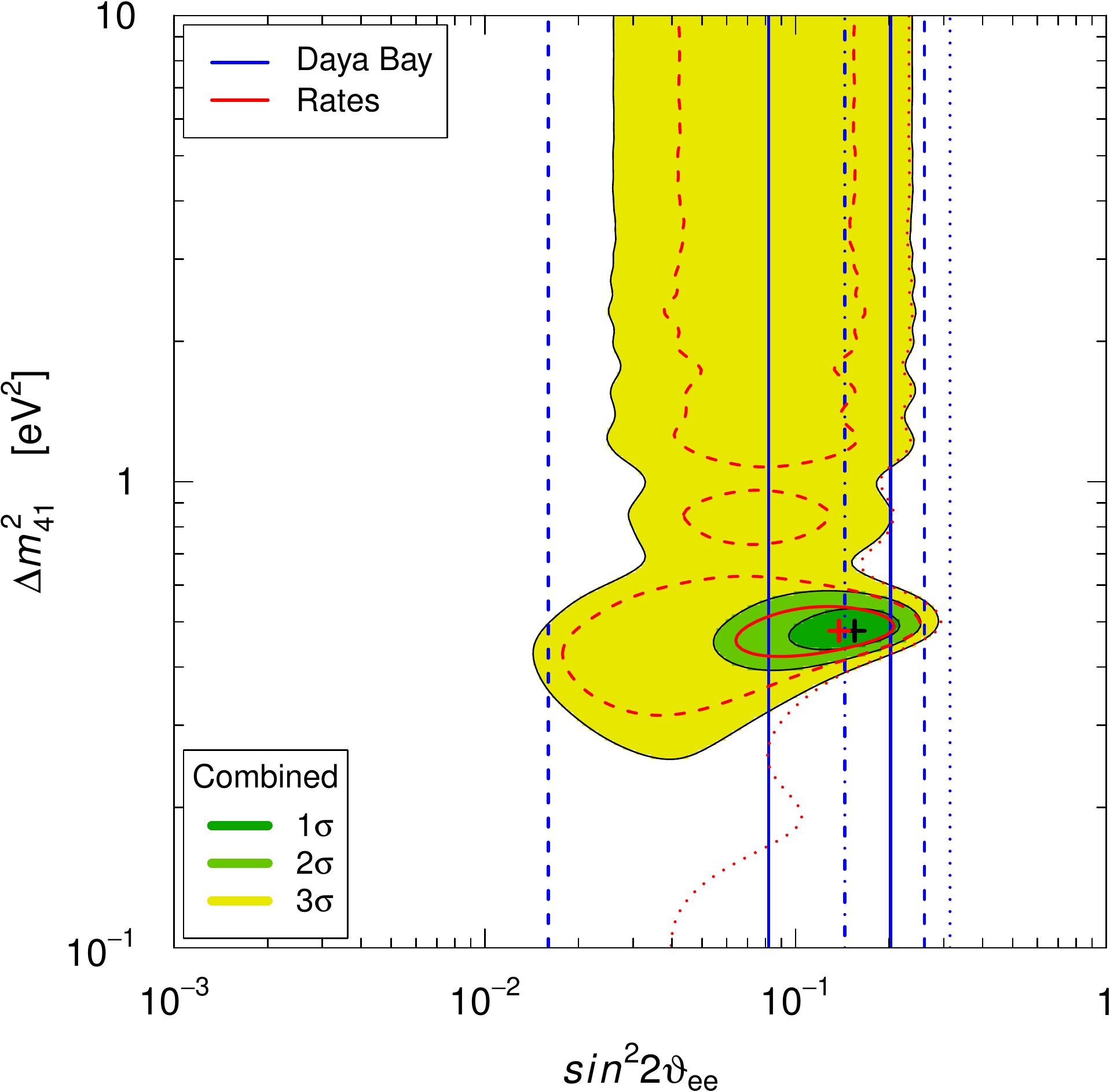}
\caption{ \label{fig:239+OSC}
Allowed regions in the $\sin^22\vartheta_{ee}$--$\Delta{m}^2_{41}$ plane
obtained
from the fit of the Daya Bay evolution data~\cite{An:2017osx} (Daya Bay),
from the fit of the reactor rates (Rates),
and from the combined fit (Combined)
with the \textbf{\ref{case:239+OSC}} model.
The best fit points are indicated by crosses,
except for the fit of the Daya Bay evolution data for which
the best fit is the vertical dash-dotted line.
For the Daya Bay and Rates fits the
$1\sigma$,
$2\sigma$, and
$3\sigma$
allowed regions are limited,
respectively,
by solid, dashed, and dotted lines.
}
\end{figure}

Figures~\ref{fig:235}, \ref{fig:235+239}, \ref{fig:OSC}, \ref{fig:235+OSC}, and \ref{fig:239+OSC}
show the allowed regions of the fitted parameters in the
\textbf{\ref{case:235}},
\textbf{\ref{case:235+239}},
\textbf{\ref{case:OSC}},
\textbf{\ref{case:235+OSC}}, and
\textbf{\ref{case:239+OSC}}
models, respectively.
In these figures,
the results of the fit of the Daya Bay evolution data are compared with those
of the fit of the reactor rates discussed in Section~\ref{sec:rea}
and
those of the combined fit discussed in Section~\ref{sec:all}.

From Fig.~\ref{fig:235}, one can see that
assuming the \textbf{\ref{case:235}} model,
the fit of the Daya Bay evolution data gives
\begin{equation}
r_{235}
=
0.927
\pm
0.022
,
\label{dby-235}
\end{equation}
which determines the ${}^{235}\text{U}$ cross section per fission to be
\begin{equation}
\sigma_{f,235}
=
6.20
\pm
0.15
.
\label{dby-csf235}
\end{equation}

In the case of the \textbf{\ref{case:235+239}} model,
Fig.~\ref{fig:235+239}
show that the Daya Bay evolution data
indicate a larger suppression of $\sigma_{f,235}$ than  $\sigma_{f,239}$,
in agreement with the results of the
analysis of the Daya Bay collaboration~\cite{An:2017osx}.
We obtained
\begin{align}
r_{235}
=
\null & \null
0.922
\pm
0.025
,
\label{dby-235+239:r35}
\\
r_{239}
=
\null & \null
0.974
\pm
0.046
,
\label{dby-235+239:r39}
\end{align}
which imply
\begin{align}
\sigma_{f,235}
=
\null & \null
6.17
\pm
0.16
,
\label{dby-235+239:c35}
\\
\sigma_{f,239}
=
\null & \null
4.29
\pm
0.20
.
\label{dby-235+239:c39}
\end{align}
These results are compatible with those obtained by the Daya Bay collaboration~\cite{An:2017osx},
taking into account of the different assumptions on the uncertainties of
$\sigma_{f,238}$ and $\sigma_{f,241}$
(10\% in the calculation of the Daya Bay collaboration
and
the Saclay+Huber theoretical values~\cite{Mueller:2011nm,Mention:2011rk,Huber:2011wv}
8.15\% and 2.60\%
in our calculation).

The vertical lines in Fig.~\ref{fig:OSC}
show the bounds on
$\sin^22\vartheta_{ee} = 2 ( 1 - P_{ee} )$
obtained in the
\textbf{\ref{case:OSC}}
analysis of the Daya Bay evolution data,
in which oscillations are averaged because of the large source-detector distance.
One can see that
\begin{equation}
\sin^22\vartheta_{ee}
=
0.12
\pm
0.06
,
\label{dby-OSC:see}
\end{equation}
and there is no lower bound at $2\sigma$,
because oscillations are favored over the no-oscillation case only at the
$1.9\sigma$ level.

Figure~\ref{fig:235+OSC} shows that the variation of
$r_{235}$
in the \textbf{\ref{case:235+OSC}}
model causes a shift of the allowed region for
$\sin^22\vartheta_{ee}$
towards lower values
with respect to Fig.~\ref{fig:OSC}
obtained with the
\textbf{\ref{case:OSC}}.
In Fig.~\ref{fig:235+OSC}
there is no lower bound at $1\sigma$,
because oscillations are favored over the no-oscillation case only at
$0.4\sigma$.
This is due to the preference for values of $r_{235}$
smaller than one,
as shown by the best-fit value in Tab.~\ref{tab:dby}.

On the other hand,
in Fig.~\ref{fig:239+OSC} corresponding to the \textbf{\ref{case:239+OSC}}
there is a shift of the allowed region for
$\sin^22\vartheta_{ee}$
towards larger values
with respect to Fig.~\ref{fig:OSC}
obtained with the
\textbf{\ref{case:OSC}},
because $P_{ee}$ is smaller in order to compensate the increase of
$\sigma_{f,a}^{\text{th}}$ due to
$r_{239}>1$.

\section{Previous reactor rates}
\label{sec:rea}

\begin{table}[!b]
\centering
\begin{tabular}[t]{c}
\\
\hline
$\chi^{2}_{\text{min}}$\\
NDF\\
GoF\\
$\Delta{m}^2_{41}$\\
$\sin^22\vartheta_{ee}$\\
$r_{235}$\\
$r_{239}$\\
%\hline%
\end{tabular}%
\begin{tabular}[t]{c}
\textbf{\ref{case:235}}\\
\hline
$20.7$\\
$25$\\
$71\%$\\
$-$\\
$-$\\
$0.939$\\
$-$\\
%\hline%
\end{tabular}%
\begin{tabular}[t]{c}
\textbf{\ref{case:235+239}}\\
\hline
$17.7$\\
$24$\\
$82\%$\\
$-$\\
$-$\\
$0.950$\\
$0.873$\\
%\hline%
\end{tabular}%
\begin{tabular}[t]{c}
\textbf{\ref{case:OSC}}\\
\hline
$12.8$\\
$24$\\
$100\%$\\
$0.48$\\
$0.13$\\
$-$\\
$-$\\
%\hline%
\end{tabular}%
\begin{tabular}[t]{c}
\textbf{\ref{case:235+OSC}}\\
\hline
$12.6$\\
$23$\\
$100\%$\\
$0.48$\\
$0.15$\\
$1.025$\\
$-$\\
%\hline%
\end{tabular}%
\begin{tabular}[t]{c}
\textbf{\ref{case:239+OSC}}\\
\hline
$12.7$\\
$23$\\
$100\%$\\
$0.48$\\
$0.14$\\
$-$\\
$1.036$\\
%\hline%
\end{tabular}%
\caption{ \label{tab:rea}
Fits of the reactor rates in Table~1 of Ref.~\cite{Gariazzo:2017fdh}.
}
\end{table}

\begin{figure}[!t]
\centering
\includegraphics*[width=\linewidth]{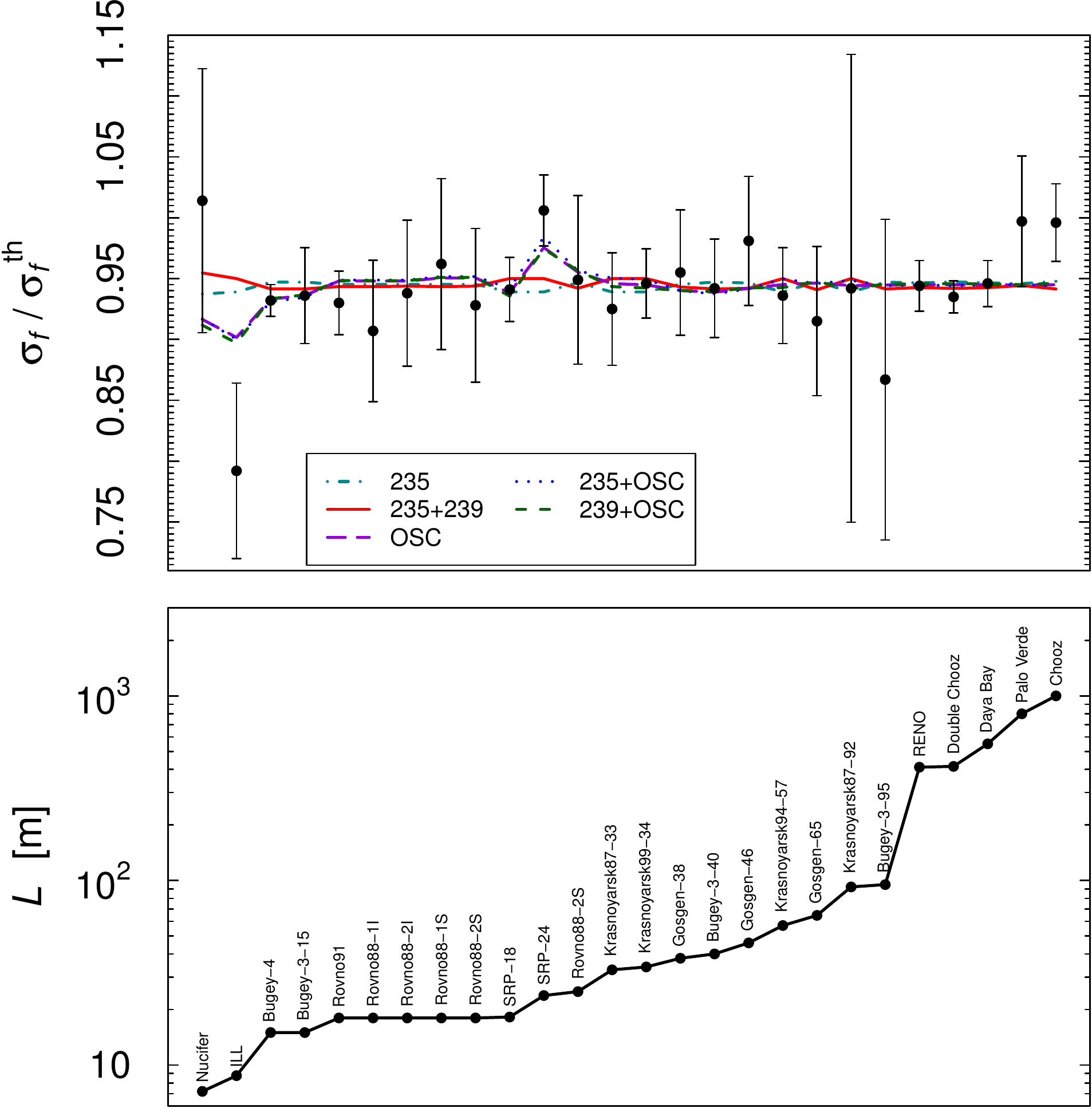}
\caption{ \label{fig:rea-evo-len}
The top panels show the
fits of the reactor rates in Table~1 of Ref.~\cite{Gariazzo:2017fdh}.
The data are ordered by increasing values of the source-detector distance $L$,
shown in the bottom panel.
The error bars show to the experimental uncertainties.
}
\end{figure}

\begin{figure}[!t]
\centering
\includegraphics*[width=\linewidth]{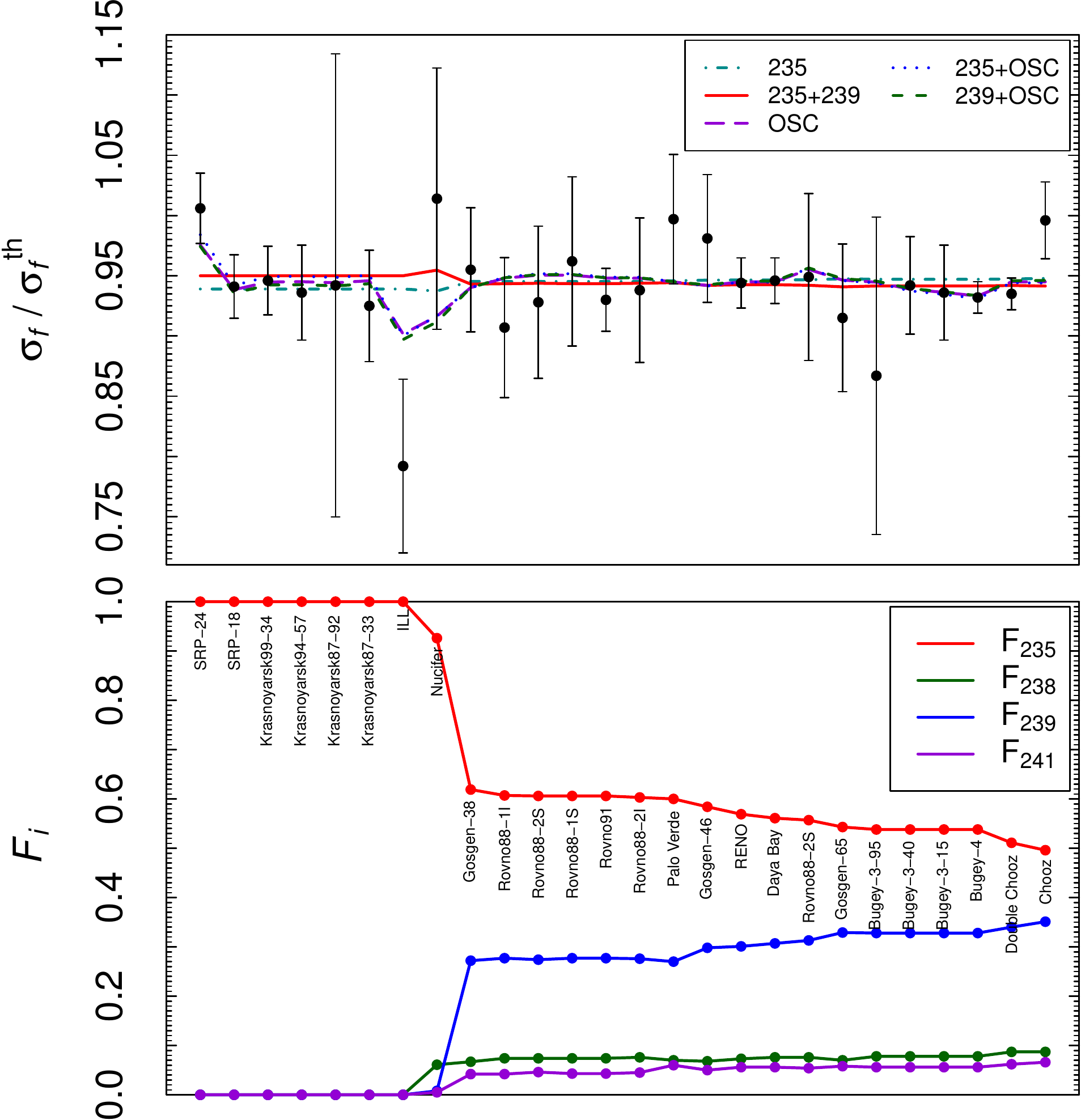}
\caption{ \label{fig:rea-evo-f35}
The top panels show the
fits of the reactor rates in Table~1 of Ref.~\cite{Gariazzo:2017fdh}.
The data are ordered by decreasing values of $F_{235}$,
shown in the bottom panel.
The error bars show to the experimental uncertainties.
}
\end{figure}

In this section we consider the reactor antineutrino data which were available
before the release of the Daya Bay fuel evolution data in Ref.~\cite{An:2017osx}.
We use the data listed in Table~1 of Ref.~\cite{Gariazzo:2017fdh}
of the following experiments:
Bugey-4~\cite{Declais:1994ma},
Rovno91~\cite{Kuvshinnikov:1990ry},
Bugey-3~\cite{Declais:1995su},
Gosgen~\cite{Zacek:1986cu},
ILL~\cite{Kwon:1981ua,Hoummada:1995zz},
Krasnoyarsk87~\cite{Vidyakin:1987ue},
Krasnoyarsk94~\cite{Vidyakin:1990iz,Vidyakin:1994ut},
Rovno88~\cite{Afonin:1988gx},
SRP~\cite{Greenwood:1996pb},
Nucifer~\cite{Boireau:2015dda},
Chooz~\cite{Apollonio:2002gd},
Palo Verde~\cite{Boehm:2001ik},
Daya Bay~\cite{An:2016srz},
RENO~\cite{RENO-AAP2016},
and
Double Chooz~\cite{DoubleChooz-private-16}.
The Daya Bay data in Ref.~\cite{An:2016srz}
are relative to the average Daya Bay fuel fractions for the corresponding
detection time.

The results of the fits with the models described in Section~\ref{sec:intro}
are listed in Tab.~\ref{tab:rea}
and the fit of the data is illustrated in Figs.~\ref{fig:rea-evo-len}
and \ref{fig:rea-evo-f35}.

From Tab.~\ref{tab:rea}
one can see that all the model
have an excellent goodness-of-fit,
but
the models
\textbf{\ref{case:OSC}},
\textbf{\ref{case:235+OSC}}, and
\textbf{\ref{case:239+OSC}}
with active-sterile neutrino oscillations have a
significantly lower value of $\chi^2_{\text{min}}$.
This is due to the different source-detector distances in the experiments.
As one can see from Fig.~\ref{fig:rea-evo-len},
where the reactor data are ordered by increasing values of the source-detector distance $L$.
One can see that
active-sterile oscillations can fit better the data of the
short-baseline experiments
which have a source detector distance between about 10 and 100 m.
On the other hand,
the poor fit of the data with the
\textbf{\ref{case:235}} model
is explained by the lack of a trend Fig.~\ref{fig:rea-evo-f35},
where the reactor data are ordered by decreasing values of $F_{235}$.

The comparison of the nested models
\textbf{\ref{case:235}} and \textbf{\ref{case:235+OSC}}
give
$\Delta\chi^2_{\text{min}} = 8.1$
with two degrees of freedom.
Hence,
the $p$-value of the null hypothesis \textbf{\ref{case:235}} is
$1.7\%$
and it can be rejected in favor of the introduction of active-sterile neutrino oscillations
at
$2.4\sigma$.
As a check,
with a Monte Carlo simulation we obtained a $p$-value of
$1.3\%$,
which corresponds to
$2.5\sigma$.

The \textbf{\ref{case:235}} and \textbf{\ref{case:OSC}}
models have
$\Delta\chi^2_{\text{min}} = 7.9$
and our Monte Carlo comparison disfavors the null hypothesis \textbf{\ref{case:235}} at
$3.1\sigma$.

The \textbf{\ref{case:235+239}} and \textbf{\ref{case:OSC}}
models have
$\Delta\chi^2_{\text{min}} = 4.9$
and our Monte Carlo comparison disfavors the null hypothesis \textbf{\ref{case:235+239}} at
$2.8\sigma$.

Figure~\ref{fig:235}
shows the marginal $\Delta\chi^2 = \chi^2 - \chi^2_{\text{min}}$
for the factor $r_{235}$
obtained from the fit of the reactor rates in the \textbf{\ref{case:235}} model.
The result is
\begin{equation}
r_{235}
=
0.939
\pm
0.012
,
\label{rea-235}
\end{equation}
which gives
\begin{equation}
\sigma_{f,235}
=
6.28
\pm
0.08
.
\label{rea-csf235}
\end{equation}
This is a determination of
$\sigma_{f,235}$
with smaller uncertainty than that obtained in Eq.~(\ref{dby-csf235})
from the Daya Bay evolution data.

Figure~\ref{fig:235+239}
shows that in the case of the \textbf{\ref{case:235+239}} model the determination of
$r_{235}$ and $r_{239}$
is quite different in the analyses of the Daya Bay evolution data and the reactor rates.
In the first analysis
$r_{235}$ and $r_{239}$
are correlated,
whereas in the second analysis they are slightly anticorrelated.
Moreover,
the analysis of the reactor rates prefers a
larger value of $r_{235}$ and a smaller value of $r_{239}$
than the analysis of the Daya Bay evolution data.
The results of the analysis of the reactor rates are
\begin{align}
r_{235}
=
\null & \null
0.950
\pm
0.013
,
\label{rea-235+239:r35}
\\
r_{239}
=
\null & \null
0.873
\pm
0.064
,
\label{rea-235+239:r39}
\end{align}
which imply
\begin{align}
\sigma_{f,235}
=
\null & \null
6.36
\pm
0.09
,
\label{rea-235+239:c35}
\\
\sigma_{f,239}
=
\null & \null
3.84
\pm
0.28
.
\label{rea-235+239:c39}
\end{align}

Figure~\ref{fig:OSC}
show the allowed region in the $\sin^22\vartheta_{ee}$--$\Delta{m}^2_{41}$ plane
obtained
from the fit of the reactor rates
in the
\textbf{\ref{case:OSC}} model.
One can see that there in only one region allowed at $1\sigma$
around the best-fit point given in Tab.~\ref{tab:rea},
but the $2\sigma$ allowed regions do not have an upper bound for $\Delta{m}^2_{41}$.
The $3\sigma$ allowed region does not have a lower bound for $\sin^22\vartheta_{ee}$,
because oscillations are favored over the no-oscillation case only at the
$2.7\sigma$ level.

From a comparison of Figs.~\ref{fig:OSC}, \ref{fig:235+OSC}, and \ref{fig:239+OSC}
one can see that the variations of
$r_{235}$ and $r_{239}$
in the \textbf{\ref{case:235+OSC}} and \textbf{\ref{case:239+OSC}}
models, respectively,
have small effects on the allowed region in the $\sin^22\vartheta_{ee}$--$\Delta{m}^2_{41}$ plane,
in agreement with the best-fit values close to one of
$r_{235}$ and $r_{239}$
in Tab.~\ref{tab:rea}.

\section{Combined analysis}
\label{sec:all}

\begin{table}[!b]
\centering
\begin{tabular}[t]{c}
\\
\hline
$\chi^{2}_{\text{min}}$\\
NDF\\
GoF\\
$\Delta{m}^2_{41}$\\
$\sin^22\vartheta_{ee}$\\
$r_{235}$\\
$r_{239}$\\
%\hline%
\end{tabular}%
\begin{tabular}[t]{c}
\textbf{\ref{case:235}}\\
\hline
$25.3$\\
$32$\\
$79\%$\\
$-$\\
$-$\\
$0.934$\\
$-$\\
%\hline%
\end{tabular}%
\begin{tabular}[t]{c}
\textbf{\ref{case:235+239}}\\
\hline
$24.8$\\
$31$\\
$78\%$\\
$-$\\
$-$\\
$0.934$\\
$0.970$\\
%\hline%
\end{tabular}%
\begin{tabular}[t]{c}
\textbf{\ref{case:OSC}}\\
\hline
$23.0$\\
$31$\\
$85\%$\\
$0.48$\\
$0.14$\\
$-$\\
$-$\\
%\hline%
\end{tabular}%
\begin{tabular}[t]{c}
\textbf{\ref{case:235+OSC}}\\
\hline
$20.2$\\
$30$\\
$91\%$\\
$0.48$\\
$0.11$\\
$0.987$\\
$-$\\
%\hline%
\end{tabular}%
\begin{tabular}[t]{c}
\textbf{\ref{case:239+OSC}}\\
\hline
$17.5$\\
$30$\\
$100\%$\\
$0.48$\\
$0.15$\\
$-$\\
$1.099$\\
%\hline%
\end{tabular}%
\caption{ \label{tab:all}
Fits of the reactor rates in Table~1 of Ref.~\cite{Gariazzo:2017fdh} (without the 2016 Daya Bay rate)
and
the 2017 Daya Bay evolution data~\cite{An:2017osx}.
}
\end{table}

In this section we present the results of the combined fits of the reactor rates in Table~1 of Ref.~\cite{Gariazzo:2017fdh} (without the 2016 Daya Bay rate)
and
the 2017 Daya Bay evolution data~\cite{An:2017osx}.

The results of the fits with the models described in Section~\ref{sec:intro}
are listed in Tab.~\ref{tab:all}.

From Tab.~\ref{tab:all}
one can see that all the models
have an excellent goodness-of-fit.
The \textbf{\ref{case:OSC}} model
has a better goodness-of-fit than the \textbf{\ref{case:235}} model.
There is little improvement of the goodness-of-fit
from the \textbf{\ref{case:235}} model
to the \textbf{\ref{case:235+239}} model,
whereas the goodness-of-fit
improves significantly in the \textbf{\ref{case:235+OSC}} model
and especially in the
\textbf{\ref{case:239+OSC}} model.

The comparison of the nested models
\textbf{\ref{case:235}} and \textbf{\ref{case:235+OSC}}
give
$\Delta\chi^2 = 5.1$
with two degrees of freedom.
Hence,
the $p$-value of the null hypothesis \textbf{\ref{case:235}} is
$7.8\%$
and it can be rejected in favor of the introduction of active-sterile neutrino oscillations
only at
$1.8\sigma$.
As a check,
with a Monte Carlo simulation we obtained a $p$-value of
$5.1\%$,
which corresponds to
$1.9\sigma$.

The \textbf{\ref{case:235}} and \textbf{\ref{case:235+239}} models
have
$\Delta\chi^2_{\text{min}} = 2.3$
and
$1.8$
with respect to the \textbf{\ref{case:OSC}} model
and our Monte Carlo comparison disfavors them at
$1.7\sigma$
and
$2.2\sigma$,
respectively.

The
\textbf{\ref{case:235}},
\textbf{\ref{case:235+239}},
\textbf{\ref{case:OSC}}, and
\textbf{\ref{case:235+OSC}}
models
have
$\Delta\chi^2_{\text{min}} = 7.8$,
$7.3$,
$5.5$, and
$2.7$
with respect to the \textbf{\ref{case:239+OSC}} model
and our Monte Carlo comparison disfavors them at
$4.2\sigma$,
$2.9\sigma$,
$2.4\sigma$, and
$3.5\sigma$,
respectively.

From Fig.~\ref{fig:235}
one can see that
in the \textbf{\ref{case:235}} model
the combined fit indicates a value of $r_{235}$
intermediate between those obtained
from the analyzes of the Daya Bay evolution data and the reactor rates.
The result is
\begin{equation}
r_{235}
=
0.934
\pm
0.010
,
\label{all-235}
\end{equation}
which gives
\begin{equation}
\sigma_{f,235}
=
6.25
\pm
0.07
.
\label{all-csf235}
\end{equation}
This is a determination of
$\sigma_{f,235}$
with smaller uncertainty than that obtained in Eq.~(\ref{dby-csf235})
from the Daya Bay evolution data
and that obtained in Eq.~(\ref{rea-csf235})
from the reactor rates.

Figure~\ref{fig:235+239}
shows that in the case of the \textbf{\ref{case:235+239}} model the determination of
$r_{235}$ and $r_{239}$
from the combined fit improves the uncertainties of the two parameters
with respect to those obtained from the separate analyses of
the Daya Bay evolution data and the reactor rates
The results are
\begin{align}
r_{235}
=
\null & \null
0.934
\pm
0.009
,
\label{all-235+239:r35}
\\
r_{239}
=
\null & \null
0.970
\pm
0.032
,
\label{all-235+239:r39}
\end{align}
which give
\begin{align}
\sigma_{f,235}
=
\null & \null
6.25
\pm
0.06
,
\label{all-235+239:c35}
\\
\sigma_{f,239}
=
\null & \null
4.27
\pm
0.14
.
\label{all-235+239:c39}
\end{align}
Within the uncertainties,
these results are compatible with those obtained in Ref.~\cite{Giunti:2017nww}
with different assumptions on the uncertainties of
$\sigma_{f,238}$ and $\sigma_{f,241}$.
Note that here we performed a full analysis of the Daya Bay evolution data
using the complete information available in the Supplemental Material
of Ref.~\cite{An:2017osx}
whereas in Ref.~\cite{Giunti:2017nww}
the Daya Bay evolution data have been taken into account
with a Gaussian approximation of the $\chi^2$ distribution
in Fig.~3 of Ref.~\cite{An:2017osx}.

Figure~\ref{fig:OSC}
show the allowed region in the $\sin^22\vartheta_{ee}$--$\Delta{m}^2_{41}$ plane
in the
\textbf{\ref{case:OSC}} model.
The allowed regions are smaller than those obtained
from the fit of the reactor rates and there is a $3\sigma$ lower bound for $\sin^22\vartheta_{ee}$,
because oscillations are favored over the no-oscillation case at the
$3.1\sigma$ level.
However, there is no upper bound for $\Delta{m}^2_{41}$ at $2\sigma$,
because at that confidence level the data can be fitted with an averaged
oscillation probability
which does not depend on the source-detector distance.

Comparing Figs.~\ref{fig:OSC} and \ref{fig:235+OSC},
one can see that the variation of
$r_{235}$
in the \textbf{\ref{case:235+OSC}}
enlarges the allowed regions towards lower values of
$\sin^22\vartheta_{ee}$
and there is no lower bound for $\sin^22\vartheta_{ee}$ at $2\sigma$,
because oscillations are favored over the no-oscillation case only at
$1.4\sigma$.
This is due to the preference for values of $r_{235}$
smaller than one,
as shown by the best-fit value in Tab.~\ref{tab:all}.

Figure~\ref{fig:239+OSC}
shows that the best-fitting model
\textbf{\ref{case:239+OSC}}
gives the strongest indication in favor of oscillations,
which are favored over the no-oscillation case at
$3.0\sigma$.
This is due to the preference for values of $r_{239}$
larger than one,
as shown by the best-fit value in Tab.~\ref{tab:all}.

\section{Conclusions}
\label{sec:conclusions}

In this paper we analyzed the Daya Bay evolution data~\cite{An:2017osx}
in the
\textbf{\ref{case:235}},
\textbf{\ref{case:235+239}},
\textbf{\ref{case:OSC}},
\textbf{\ref{case:235+OSC}}, and
\textbf{\ref{case:239+OSC}}
models described in Section~\ref{sec:intro},
which allow to compare the fits of the data under the hypotheses of
variations of the
${}^{235}\text{U}$ and ${}^{239}\text{Pu}$
reactor antineutrino fluxes
with respect to
the Saclay+Huber theoretical value~\cite{Mueller:2011nm,Mention:2011rk,Huber:2011wv}
and
short-baseline active-sterile neutrino oscillations,
taking into account the theoretical uncertainties of the reactor antineutrino fluxes.
We found that the best explanation of
the Daya Bay evolution data
is the \textbf{\ref{case:235}} model with a variation of the
${}^{235}\text{U}$
flux
with respect to
the Saclay+Huber theoretical value~\cite{Mueller:2011nm,Mention:2011rk,Huber:2011wv}.
Comparing the \textbf{\ref{case:OSC}} model of
active-sterile neutrino oscillations
with
the \textbf{\ref{case:235}} model,
we found that
it is disfavored at $2.6\sigma$.

We also compared the
\textbf{\ref{case:OSC}} model with the
\textbf{\ref{case:235+239}} model which allows
independent variations of the
${}^{235}\text{U}$ and ${}^{239}\text{Pu}$
fluxes with respect to
Saclay+Huber theoretical values~\cite{Mueller:2011nm,Mention:2011rk,Huber:2011wv}.
We found that the \textbf{\ref{case:OSC}} model is
disfavored at $2.5\sigma$.
This result is slightly less stringent
than the $2.8\sigma$ obtained by the Daya Bay collaboration~\cite{An:2017osx}
without considering the theoretical uncertainties.

The Daya Bay evolution data can also be fitted well with the
\textbf{\ref{case:235+OSC}}
model,
with a suppression of the ${}^{235}\text{U}$ flux and neutrino oscillations,
or with the
\textbf{\ref{case:239+OSC}}
model,
with an enhancement of the ${}^{239}\text{Pu}$ flux and relatively large neutrino oscillations.

We also performed a similar analysis of
the reactor antineutrino data which were available
before the release of the Daya Bay fuel evolution data in Ref.~\cite{An:2017osx}.
In this case, we found that
the best explanation of the data is the
\textbf{\ref{case:OSC}} model with
active-sterile neutrino oscillations,
which depend on the source-detector distance
and fit the rates measured by reactor experiments with a source-detector distance
between about 10 and 100 m better than the distance-independent suppression of the
reactor antineutrino flux
given by suppressions of the
${}^{235}\text{U}$ and ${}^{239}\text{Pu}$
fluxes.
In this case,
the \textbf{\ref{case:235}} model with a suppression of the
${}^{235}\text{U}$
flux only is disfavored at
$3.1\sigma$
and the \textbf{\ref{case:235+239}} model with independent suppressions of the
${}^{235}\text{U}$ and ${}^{239}\text{Pu}$
fluxes is disfavored at
$2.8\sigma$.
As with the fit of the Daya Bay evolution data, composite models including both
variations of the
${}^{235}\text{U}$ or ${}^{239}\text{Pu}$
fluxes
and active-sterile oscillations provide good fits to the global reactor rate data.

Finally, we performed combined fits of the Daya Bay evolution data and the other reactor rates
and we found that all the considered models fit well the data.
The \textbf{\ref{case:OSC}} model has a better goodness-of-fit
than the
\textbf{\ref{case:235}} and \textbf{\ref{case:235+239}} models,
which are almost equivalent.
We obtained better fits of the data with
the composite \textbf{\ref{case:235+OSC}} and \textbf{\ref{case:239+OSC}} models.
In particular,
the best-fit model is
\textbf{\ref{case:239+OSC}},
with an increase of the
${}^{239}\text{Pu}$ flux
with respect to
the Saclay+Huber theoretical value~\cite{Mueller:2011nm,Mention:2011rk,Huber:2011wv}
and relatively large
active-sterile neutrino oscillations.

In conclusion,
although the recent Daya Bay evolution data~\cite{An:2017osx}
disfavor short-baseline active-sterile neutrino oscillations
over a suppression of the ${}^{235}\text{U}$ reactor antineutrino flux
or independent suppressions of the
${}^{235}\text{U}$ and ${}^{239}\text{Pu}$
fluxes,
the result is reversed in the analysis of the other available reactor antineutrino data.
Both sets of data are individually well-fitted by composite models
with
variations of the
${}^{235}\text{U}$ or ${}^{239}\text{Pu}$
fluxes
and active-sterile neutrino oscillations.
The combined data set indicates a preference for
the composite models and,
in particular,
the best fit is obtained with the
\textbf{\ref{case:239+OSC}} model,
through an enhancement of the
${}^{239}\text{Pu}$
flux
and relatively large oscillations.
However,
while these combined fits suggest a preference for models including sterile neutrinos,
the significant uncertainties in the reactor rate measurements and
the high goodness-of-fits observed for models both with and without sterile neutrinos
make it clear that
the search for the explanation of the reactor antineutrino anomaly~\cite{Mention:2011rk}
still remains open.
We hope that it will be solved soon by the new short-baseline
reactor neutrino experiments
which will measure the reactor antineutrino flux from reactors with different fuel compositions:
highly enriched ${}^{235}\text{U}$ research reactors for
PROSPECT \cite{Ashenfelter:2015uxt},
SoLid \cite{Ryder:2015sma}, and
STEREO \cite{Helaine:2016bmc},
and commercial reactors with mixed fuel compositions for
DANSS \cite{Alekseev:2016llm} and
Neutrino-4 \cite{Serebrov:2017nxa}.

\section*{Acknowledgment}

We would like to thank the Daya Bay collaboration for useful
discussions and information on the
Daya Bay evolution data.
The work of X.P. Ji was supported by the National Natural Science Foundation of China (Grants No. 11235006 and No. 11475093)
and by the CAS Center for Excellence in Particle Physics (CCEPP).
The work of Y.F. Li was supported in part by the National Natural Science Foundation of China under Grant Nos. 11305193 and 11135009, by the Strategic Priority Research Program of the Chinese Academy of Sciences under Grant No. XDA10010100, and by CCEPP.
The work of B.R. Littlejohn was partially supported by the DOE Office of Science, under award No. DE-SC0008347.

%\bibliographystyle{apsrev4-1}
%\bibliography{evol}

\begin{thebibliography}{29}%
\makeatletter
\providecommand \@ifxundefined [1]{%
\@ifx{#1\undefined}
}%
\providecommand \@ifnum [1]{%
\ifnum #1\expandafter \@firstoftwo
\else \expandafter \@secondoftwo
\fi
}%
\providecommand \@ifx [1]{%
\ifx #1\expandafter \@firstoftwo
\else \expandafter \@secondoftwo
\fi
}%
\providecommand \natexlab [1]{#1}%
\providecommand \enquote [1]{``#1''}%
\providecommand \bibnamefont [1]{#1}%
\providecommand \bibfnamefont [1]{#1}%
\providecommand \citenamefont [1]{#1}%
\providecommand \href@noop [0]{\@secondoftwo}%
\providecommand \href [0]{\begingroup \@sanitize@url \@href}%
\providecommand \@href[1]{\@@startlink{#1}\@@href}%
\providecommand \@@href[1]{\endgroup#1\@@endlink}%
\providecommand \@sanitize@url [0]{\catcode `\\12\catcode `\$12\catcode
`\&12\catcode `\#12\catcode `\^12\catcode `\_12\catcode `\%12\relax}%
\providecommand \@@startlink[1]{}%
\providecommand \@@endlink[0]{}%
\providecommand \url [0]{\begingroup\@sanitize@url \@url }%
\providecommand \@url [1]{\endgroup\@href {#1}{\urlprefix }}%
\providecommand \urlprefix [0]{URL }%
\providecommand \Eprint [0]{\href }%
\providecommand \doibase [0]{http://dx.doi.org/}%
\providecommand \selectlanguage [0]{\@gobble}%
\providecommand \bibinfo [0]{\@secondoftwo}%
\providecommand \bibfield [0]{\@secondoftwo}%
\providecommand \translation [1]{[#1]}%
\providecommand \BibitemOpen [0]{}%
\providecommand \bibitemStop [0]{}%
\providecommand \bibitemNoStop [0]{.\EOS\space}%
\providecommand \EOS [0]{\spacefactor3000\relax}%
\providecommand \BibitemShut [1]{\csname bibitem#1\endcsname}%
\let\auto@bib@innerbib\@empty
%</preamble>
\bibitem [{\citenamefont {An}\ \emph {et~al.}(2017{\natexlab{a}})\citenamefont
{An} \emph {et~al.}}]{An:2017osx}%
\BibitemOpen
\bibfield {author} {\bibinfo {author} {\bibfnamefont {F.~P.}\ \bibnamefont
{An}} \emph {et~al.} (\bibinfo {collaboration} {Daya Bay}),\ }\href@noop {}
{\bibfield {journal} {\bibinfo {journal} {Phys.Rev.Lett.}\ }\textbf
{\bibinfo {volume} {118}},\ \bibinfo {pages} {251801} (\bibinfo {year}
{2017}{\natexlab{a}})},\ \Eprint {http://arxiv.org/abs/arXiv:1704.01082}
{arXiv:1704.01082 [physics]} \BibitemShut {NoStop}%
\bibitem [{\citenamefont {Mueller}\ \emph {et~al.}(2011)\citenamefont {Mueller}
\emph {et~al.}}]{Mueller:2011nm}%
\BibitemOpen
\bibfield {author} {\bibinfo {author} {\bibfnamefont {T.~A.}\ \bibnamefont
{Mueller}} \emph {et~al.},\ }\href@noop {} {\bibfield {journal} {\bibinfo
{journal} {Phys. Rev.}\ }\textbf {\bibinfo {volume} {C83}},\ \bibinfo {pages}
{054615} (\bibinfo {year} {2011})},\ \Eprint
{http://arxiv.org/abs/arXiv:1101.2663} {arXiv:1101.2663 [hep-ex]}
\BibitemShut {NoStop}%
\bibitem [{\citenamefont {Mention}\ \emph {et~al.}(2011)\citenamefont {Mention}
\emph {et~al.}}]{Mention:2011rk}%
\BibitemOpen
\bibfield {author} {\bibinfo {author} {\bibfnamefont {G.}~\bibnamefont
{Mention}} \emph {et~al.},\ }\href@noop {} {\bibfield {journal} {\bibinfo
{journal} {Phys. Rev.}\ }\textbf {\bibinfo {volume} {D83}},\ \bibinfo {pages}
{073006} (\bibinfo {year} {2011})},\ \Eprint
{http://arxiv.org/abs/arXiv:1101.2755} {arXiv:1101.2755 [hep-ex]}
\BibitemShut {NoStop}%
\bibitem [{\citenamefont {Huber}(2011)}]{Huber:2011wv}%
\BibitemOpen
\bibfield {author} {\bibinfo {author} {\bibfnamefont {P.}~\bibnamefont
{Huber}},\ }\href@noop {} {\bibfield {journal} {\bibinfo {journal} {Phys.
Rev.}\ }\textbf {\bibinfo {volume} {C84}},\ \bibinfo {pages} {024617}
(\bibinfo {year} {2011})},\ \Eprint {http://arxiv.org/abs/arXiv:1106.0687}
{arXiv:1106.0687 [hep-ph]} \BibitemShut {NoStop}%
\bibitem [{\citenamefont {Gariazzo}\ \emph {et~al.}(2017)\citenamefont
{Gariazzo}, \citenamefont {Giunti}, \citenamefont {Laveder},\ and\
\citenamefont {Li}}]{Gariazzo:2017fdh}%
\BibitemOpen
\bibfield {author} {\bibinfo {author} {\bibfnamefont {S.}~\bibnamefont
{Gariazzo}}, \bibinfo {author} {\bibfnamefont {C.}~\bibnamefont {Giunti}},
\bibinfo {author} {\bibfnamefont {M.}~\bibnamefont {Laveder}}, \ and\
\bibinfo {author} {\bibfnamefont {Y.}~\bibnamefont {Li}},\ }\href@noop {}
{\bibfield {journal} {\bibinfo {journal} {JHEP}\ }\textbf {\bibinfo
{volume} {1706}},\ \bibinfo {pages} {135} (\bibinfo {year} {2017})},\ \Eprint
{http://arxiv.org/abs/arXiv:1703.00860} {arXiv:1703.00860 [hep-ph]}
\BibitemShut {NoStop}%
\bibitem [{\citenamefont {Gariazzo}\ \emph {et~al.}(2016)\citenamefont
{Gariazzo}, \citenamefont {Giunti}, \citenamefont {Laveder}, \citenamefont
{Li},\ and\ \citenamefont {Zavanin}}]{Gariazzo:2015rra}%
\BibitemOpen
\bibfield {author} {\bibinfo {author} {\bibfnamefont {S.}~\bibnamefont
{Gariazzo}}, \bibinfo {author} {\bibfnamefont {C.}~\bibnamefont {Giunti}},
\bibinfo {author} {\bibfnamefont {M.}~\bibnamefont {Laveder}}, \bibinfo
{author} {\bibfnamefont {Y.}~\bibnamefont {Li}}, \ and\ \bibinfo {author}
{\bibfnamefont {E.}~\bibnamefont {Zavanin}},\ }\href@noop {} {\bibfield
{journal} {\bibinfo {journal} {J. Phys.}\ }\textbf {\bibinfo {volume}
{G43}},\ \bibinfo {pages} {033001} (\bibinfo {year} {2016})},\ \Eprint
{http://arxiv.org/abs/arXiv:1507.08204} {arXiv:1507.08204 [hep-ph]}
\BibitemShut {NoStop}%
\bibitem [{\citenamefont {Declais}\ \emph {et~al.}(1994)\citenamefont {Declais}
\emph {et~al.}}]{Declais:1994ma}%
\BibitemOpen
\bibfield {author} {\bibinfo {author} {\bibfnamefont {Y.}~\bibnamefont
{Declais}} \emph {et~al.} (\bibinfo {collaboration} {Bugey}),\ }\href@noop {}
{\bibfield {journal} {\bibinfo {journal} {Phys. Lett.}\ }\textbf {\bibinfo
{volume} {B338}},\ \bibinfo {pages} {383} (\bibinfo {year}
{1994})}\BibitemShut {NoStop}%
\bibitem [{\citenamefont {Kuvshinnikov}\ \emph {et~al.}(1991)\citenamefont
{Kuvshinnikov}, \citenamefont {Mikaelyan}, \citenamefont {Nikolaev},
\citenamefont {Skorokhvatov},\ and\ \citenamefont
{Etenko}}]{Kuvshinnikov:1990ry}%
\BibitemOpen
\bibfield {author} {\bibinfo {author} {\bibfnamefont {A.}~\bibnamefont
{Kuvshinnikov}}, \bibinfo {author} {\bibfnamefont {L.}~\bibnamefont
{Mikaelyan}}, \bibinfo {author} {\bibfnamefont {S.}~\bibnamefont {Nikolaev}},
\bibinfo {author} {\bibfnamefont {M.}~\bibnamefont {Skorokhvatov}}, \ and\
\bibinfo {author} {\bibfnamefont {A.}~\bibnamefont {Etenko}},\ }\href@noop {}
{\bibfield {journal} {\bibinfo {journal} {JETP Lett.}\ }\textbf {\bibinfo
{volume} {54}},\ \bibinfo {pages} {253} (\bibinfo {year} {1991})}\BibitemShut
{NoStop}%
\bibitem [{\citenamefont {Achkar}\ \emph {et~al.}(1995)\citenamefont {Achkar}
\emph {et~al.}}]{Declais:1995su}%
\BibitemOpen
\bibfield {author} {\bibinfo {author} {\bibfnamefont {B.}~\bibnamefont
{Achkar}} \emph {et~al.} (\bibinfo {collaboration} {Bugey}),\ }\href@noop {}
{\bibfield {journal} {\bibinfo {journal} {Nucl. Phys.}\ }\textbf {\bibinfo
{volume} {B434}},\ \bibinfo {pages} {503} (\bibinfo {year}
{1995})}\BibitemShut {NoStop}%
\bibitem [{\citenamefont {Zacek}\ \emph {et~al.}(1986)\citenamefont {Zacek}
\emph {et~al.}}]{Zacek:1986cu}%
\BibitemOpen
\bibfield {author} {\bibinfo {author} {\bibfnamefont {G.}~\bibnamefont
{Zacek}} \emph {et~al.} (\bibinfo {collaboration} {CalTech-SIN-TUM}),\
}\href@noop {} {\bibfield {journal} {\bibinfo {journal} {Phys. Rev.}\
}\textbf {\bibinfo {volume} {D34}},\ \bibinfo {pages} {2621} (\bibinfo {year}
{1986})}\BibitemShut {NoStop}%
\bibitem [{\citenamefont {Kwon}\ \emph {et~al.}(1981)\citenamefont {Kwon} \emph
{et~al.}}]{Kwon:1981ua}%
\BibitemOpen
\bibfield {author} {\bibinfo {author} {\bibfnamefont {H.}~\bibnamefont
{Kwon}} \emph {et~al.},\ }\href@noop {} {\bibfield {journal} {\bibinfo
{journal} {Phys. Rev.}\ }\textbf {\bibinfo {volume} {D24}},\ \bibinfo {pages}
{1097} (\bibinfo {year} {1981})}\BibitemShut {NoStop}%
\bibitem [{\citenamefont {Hoummada}\ \emph {et~al.}(1995)\citenamefont
{Hoummada}, \citenamefont {Lazrak~Mikou}, \citenamefont {Bagieu},
\citenamefont {Cavaignac},\ and\ \citenamefont
{Holm~Koang}}]{Hoummada:1995zz}%
\BibitemOpen
\bibfield {author} {\bibinfo {author} {\bibfnamefont {A.}~\bibnamefont
{Hoummada}}, \bibinfo {author} {\bibfnamefont {S.}~\bibnamefont
{Lazrak~Mikou}}, \bibinfo {author} {\bibfnamefont {G.}~\bibnamefont
{Bagieu}}, \bibinfo {author} {\bibfnamefont {J.}~\bibnamefont {Cavaignac}}, \
and\ \bibinfo {author} {\bibfnamefont {D.}~\bibnamefont {Holm~Koang}},\
}\href@noop {} {\bibfield {journal} {\bibinfo {journal} {Applied Radiation
and Isotopes}\ }\textbf {\bibinfo {volume} {46}},\ \bibinfo {pages} {449}
(\bibinfo {year} {1995})}\BibitemShut {NoStop}%
\bibitem [{\citenamefont {Vidyakin}\ \emph {et~al.}(1987)\citenamefont
{Vidyakin} \emph {et~al.}}]{Vidyakin:1987ue}%
\BibitemOpen
\bibfield {author} {\bibinfo {author} {\bibfnamefont {G.~S.}\ \bibnamefont
{Vidyakin}} \emph {et~al.} (\bibinfo {collaboration} {Krasnoyarsk}),\
}\href@noop {} {\bibfield {journal} {\bibinfo {journal} {Sov. Phys. JETP}\
}\textbf {\bibinfo {volume} {66}},\ \bibinfo {pages} {243} (\bibinfo {year}
{1987})}\BibitemShut {NoStop}%
\bibitem [{\citenamefont {Vidyakin}\ \emph {et~al.}(1990)\citenamefont
{Vidyakin} \emph {et~al.}}]{Vidyakin:1990iz}%
\BibitemOpen
\bibfield {author} {\bibinfo {author} {\bibfnamefont {G.~S.}\ \bibnamefont
{Vidyakin}} \emph {et~al.} (\bibinfo {collaboration} {Krasnoyarsk}),\
}\href@noop {} {\bibfield {journal} {\bibinfo {journal} {Sov. Phys. JETP}\
}\textbf {\bibinfo {volume} {71}},\ \bibinfo {pages} {424} (\bibinfo {year}
{1990})}\BibitemShut {NoStop}%
\bibitem [{\citenamefont {Vidyakin}\ \emph {et~al.}(1994)\citenamefont
{Vidyakin} \emph {et~al.}}]{Vidyakin:1994ut}%
\BibitemOpen
\bibfield {author} {\bibinfo {author} {\bibfnamefont {G.~S.}\ \bibnamefont
{Vidyakin}} \emph {et~al.} (\bibinfo {collaboration} {Krasnoyarsk}),\
}\href@noop {} {\bibfield {journal} {\bibinfo {journal} {JETP Lett.}\
}\textbf {\bibinfo {volume} {59}},\ \bibinfo {pages} {390} (\bibinfo {year}
{1994})}\BibitemShut {NoStop}%
\bibitem [{\citenamefont {Afonin}\ \emph {et~al.}(1988)\citenamefont {Afonin}
\emph {et~al.}}]{Afonin:1988gx}%
\BibitemOpen
\bibfield {author} {\bibinfo {author} {\bibfnamefont {A.~I.}\ \bibnamefont
{Afonin}} \emph {et~al.},\ }\href@noop {} {\bibfield {journal} {\bibinfo
{journal} {Sov. Phys. JETP}\ }\textbf {\bibinfo {volume} {67}},\ \bibinfo
{pages} {213} (\bibinfo {year} {1988})}\BibitemShut {NoStop}%
\bibitem [{\citenamefont {Greenwood}\ \emph {et~al.}(1996)\citenamefont
{Greenwood} \emph {et~al.}}]{Greenwood:1996pb}%
\BibitemOpen
\bibfield {author} {\bibinfo {author} {\bibfnamefont {Z.~D.}\ \bibnamefont
{Greenwood}} \emph {et~al.},\ }\href@noop {} {\bibfield {journal} {\bibinfo
{journal} {Phys. Rev.}\ }\textbf {\bibinfo {volume} {D53}},\ \bibinfo {pages}
{6054} (\bibinfo {year} {1996})}\BibitemShut {NoStop}%
\bibitem [{\citenamefont {Boireau}\ \emph {et~al.}(2016)\citenamefont {Boireau}
\emph {et~al.}}]{Boireau:2015dda}%
\BibitemOpen
\bibfield {author} {\bibinfo {author} {\bibfnamefont {G.}~\bibnamefont
{Boireau}} \emph {et~al.} (\bibinfo {collaboration} {NUCIFER}),\ }\href@noop
{} {\bibfield {journal} {\bibinfo {journal} {Phys. Rev.}\ }\textbf
{\bibinfo {volume} {D93}},\ \bibinfo {pages} {112006} (\bibinfo {year}
{2016})},\ \Eprint {http://arxiv.org/abs/arXiv:1509.05610} {arXiv:1509.05610
[physics]} \BibitemShut {NoStop}%
\bibitem [{\citenamefont {Apollonio}\ \emph {et~al.}(2003)\citenamefont
{Apollonio} \emph {et~al.}}]{Apollonio:2002gd}%
\BibitemOpen
\bibfield {author} {\bibinfo {author} {\bibfnamefont {M.}~\bibnamefont
{Apollonio}} \emph {et~al.} (\bibinfo {collaboration} {CHOOZ}),\ }\href@noop
{} {\bibfield {journal} {\bibinfo {journal} {Eur. Phys. J.}\ }\textbf
{\bibinfo {volume} {C27}},\ \bibinfo {pages} {331} (\bibinfo {year}
{2003})},\ \Eprint {http://arxiv.org/abs/hep-ex/0301017} {hep-ex/0301017}
\BibitemShut {NoStop}%
\bibitem [{\citenamefont {Boehm}\ \emph {et~al.}(2001)\citenamefont {Boehm}
\emph {et~al.}}]{Boehm:2001ik}%
\BibitemOpen
\bibfield {author} {\bibinfo {author} {\bibfnamefont {F.}~\bibnamefont
{Boehm}} \emph {et~al.} (\bibinfo {collaboration} {Palo Verde}),\ }\href@noop
{} {\bibfield {journal} {\bibinfo {journal} {Phys. Rev.}\ }\textbf
{\bibinfo {volume} {D64}},\ \bibinfo {pages} {112001} (\bibinfo {year}
{2001})},\ \Eprint {http://arxiv.org/abs/hep-ex/0107009} {hep-ex/0107009}
\BibitemShut {NoStop}%
\bibitem [{\citenamefont {An}\ \emph {et~al.}(2017{\natexlab{b}})\citenamefont
{An} \emph {et~al.}}]{An:2016srz}%
\BibitemOpen
\bibfield {author} {\bibinfo {author} {\bibfnamefont {F.}~\bibnamefont {An}}
\emph {et~al.} (\bibinfo {collaboration} {Daya Bay}),\ }\href@noop {}
{\bibfield {journal} {\bibinfo {journal} {Chin.Phys.}\ }\textbf {\bibinfo
{volume} {C41}},\ \bibinfo {pages} {013002} (\bibinfo {year}
{2017}{\natexlab{b}})},\ \Eprint {http://arxiv.org/abs/arXiv:1607.05378}
{arXiv:1607.05378 [hep-ex]} \BibitemShut {NoStop}%
\bibitem [{\citenamefont {Seo}(2016)}]{RENO-AAP2016}%
\BibitemOpen
\bibfield {author} {\bibinfo {author} {\bibfnamefont {H.}~\bibnamefont
{Seo}},\ }\href@noop {} {\ (\bibinfo {year} {2016})},\ \bibinfo {note} {talk
presented at {AAP 2016, Applied Antineutrino Physics, 1-2 December 2016,
Liverpool, UK}}\BibitemShut {NoStop}%
\bibitem [{Dou()}]{DoubleChooz-private-16}%
\BibitemOpen
\href@noop {} {\ }\bibinfo {note} {{Double Chooz Collaboration, Private
Communication}}\BibitemShut {NoStop}%
\bibitem [{\citenamefont {Giunti}()}]{Giunti:2017nww}%
\BibitemOpen
\bibfield {author} {\bibinfo {author} {\bibfnamefont {C.}~\bibnamefont
{Giunti}},\ }\href@noop {} {\ }\Eprint
{http://arxiv.org/abs/arXiv:1704.02276} {arXiv:1704.02276 [hep-ph]}
\BibitemShut {NoStop}%
\bibitem [{\citenamefont {Ashenfelter}\ \emph {et~al.}(2016)\citenamefont
{Ashenfelter} \emph {et~al.}}]{Ashenfelter:2015uxt}%
\BibitemOpen
\bibfield {author} {\bibinfo {author} {\bibfnamefont {J.}~\bibnamefont
{Ashenfelter}} \emph {et~al.} (\bibinfo {collaboration} {PROSPECT}),\
}\href@noop {} {\bibfield {journal} {\bibinfo {journal} {J. Phys.}\
}\textbf {\bibinfo {volume} {G43}},\ \bibinfo {pages} {113001} (\bibinfo
{year} {2016})},\ \Eprint {http://arxiv.org/abs/arXiv:1512.02202}
{arXiv:1512.02202 [physics]} \BibitemShut {NoStop}%
\bibitem [{\citenamefont {Ryder}(2015)}]{Ryder:2015sma}%
\BibitemOpen
\bibfield {author} {\bibinfo {author} {\bibfnamefont {N.}~\bibnamefont
{Ryder}} (\bibinfo {collaboration} {SoLid}),\ }\href@noop {} {\bibfield
{journal} {\bibinfo {journal} {PoS}\ }\textbf {\bibinfo {volume}
{EPS-HEP2015}},\ \bibinfo {pages} {071} (\bibinfo {year} {2015})},\ \Eprint
{http://arxiv.org/abs/arXiv:1510.07835} {arXiv:1510.07835 [hep-ex]}
\BibitemShut {NoStop}%
\bibitem [{\citenamefont {Helaine}()}]{Helaine:2016bmc}%
\BibitemOpen
\bibfield {author} {\bibinfo {author} {\bibfnamefont {V.}~\bibnamefont
{Helaine}} (\bibinfo {collaboration} {STEREO}),\ }\href@noop {} {\ }\Eprint
{http://arxiv.org/abs/arXiv:1604.08877} {arXiv:1604.08877 [physics.ins-det]}
\BibitemShut {NoStop}%
\bibitem [{\citenamefont {Alekseev}\ \emph {et~al.}(2016)\citenamefont
{Alekseev} \emph {et~al.}}]{Alekseev:2016llm}%
\BibitemOpen
\bibfield {author} {\bibinfo {author} {\bibfnamefont {I.}~\bibnamefont
{Alekseev}} \emph {et~al.} (\bibinfo {collaboration} {DANSS}),\ }\href@noop
{} {\bibfield {journal} {\bibinfo {journal} {JINST}\ }\textbf {\bibinfo
{volume} {11}},\ \bibinfo {pages} {P11011} (\bibinfo {year} {2016})},\
\Eprint {http://arxiv.org/abs/arXiv:1606.02896} {arXiv:1606.02896 [physics]}
\BibitemShut {NoStop}%
\bibitem [{\citenamefont {Serebrov}\ \emph {et~al.}(2017)\citenamefont
{Serebrov} \emph {et~al.}}]{Serebrov:2017nxa}%
\BibitemOpen
\bibfield {author} {\bibinfo {author} {\bibfnamefont {A.~P.}\ \bibnamefont
{Serebrov}} \emph {et~al.} (\bibinfo {collaboration} {Neutrino-4}),\
}\href@noop {} {\bibfield {journal} {\bibinfo {journal} {PoS}\ }\textbf
{\bibinfo {volume} {INPC2016}},\ \bibinfo {pages} {255} (\bibinfo {year}
{2017})},\ \Eprint {http://arxiv.org/abs/arXiv:1702.00941} {arXiv:1702.00941
[physics.ins-det]} \BibitemShut {NoStop}%
\end{thebibliography}

\end{document}